\documentclass[final,5p,times,twocolumn,authoryear]{elsarticle}

\usepackage{amssymb}
\usepackage{lipsum}
\usepackage[colorlinks=true,linkcolor=blue,citecolor=blue]{hyperref}
\usepackage{tikz}
\usepackage{graphicx}
\usepackage{txfonts}
\usepackage{amsmath}
\usepackage{bbold}
\usepackage{lipsum}
\usepackage{tikz}
\usepackage{subcaption}
\usepackage[normalem]{ulem}
\usepackage{placeins}   
\usepackage{hyperref}   
\usepackage{natbib}     
\usepackage{xcolor}

\journal{High Energy Astrophysics}

\begin{document}

\begin{frontmatter}

\title{Blazar flares from plasma blobs crossing the broad-line region}

\author[1]{S. Le Bihan\corref{cor1}}
\ead{lebihan@apc.in2p3.fr}

\author[2]{A. Dmytriiev}
\author[3]{A. Zech}

\address[1]{Université Paris Cité, CNRS, Astroparticule et Cosmologie, F-75013 Paris, France}
\address[2]{Centre for Space Research, North-West University, Potchefstroom, 2520, South Africa}
\address[3]{LUX, Observatoire de Paris, Université PSL, Sorbonne Université, CNRS, 92190 Meudon, France}

\cortext[cor1]{Corresponding author}

\begin{abstract}

{The blazar 3C\,279 is well known for its rapid and large-amplitude variability. 
On 20 December 2013, the source exhibited an orphan $\gamma$-ray flare 
characterized by a flux-doubling timescale of a few hours, a very hard spectrum, 
a time-asymmetric light curve with a slow decay, and no significant optical variability.}
{We propose a new interpretation of this event based on a two-zone scenario in which a stationary emission region produces the quiescent emission, while a second zone accelerates within the broad-line region (BLR). }
{We compute the time-dependent radiative output of both zones with the \texttt{EMBLEM} code, including synchrotron, synchrotron self-Compton, and external inverse-Compton processes, as well as bulk acceleration, adiabatic expansion, and a Fokker-Planck treatment of the electron distribution. This is the first attempt to precisely model the asymmetric $\gamma$-ray flux evolution during this flare.}
{A model with a stationary region outside the dusty torus and an accelerating plasma blob reproduces the main features of the event: 
a short and intense $\gamma$-ray flare with a hard spectrum and no optical counterpart. 
The flare results from the variation of the external photon field in the blob frame as the blob crosses the BLR and reaches its terminal Lorentz factor not far from the inner radius of the BLR.}
{Bulk acceleration and the propagation of a plasma blob within the jet provide a natural mechanism for producing high-energy flares and asymmetric light curves without requiring an {\it ad hoc} time-dependent particle injection. The model predicts a delayed EUV/X-ray enhancement once the blob exits the BLR. No very-high-energy data are available for this event, but if $\gamma$-rays were emitted in this band, a delay would be expected with respect to the {\it Fermi}-LAT flare.}

\end{abstract}


\end{frontmatter}

\section{Introduction}
\label{introduction}

Flat Spectrum Radio Quasars (FSRQs) are high-luminosity Active Galactic Nuclei (AGNs) with a relativistic jet pointed toward the Earth. Jets are thought to be responsible for the broad-band emission of the FSRQ, with the synchrotron emission covering the radio band up to X-rays, and the inverse Compton (IC) emission from X-rays up to (very-)high-energy $\gamma$-rays. The IC is thought to be caused by the upscattering of photons coming from the synchrotron emission (synchrotron-self-Compton, SSC) and by the upscattering of photons from an external radiation field (external Inverse Compton, EIC), the source of which can be the accretion disk, the broad line region (BLR) or the dusty torus of the AGN. The EIC emission is thought to dominate in FSRQs because of their high-luminosity accretion disk and thus strong external photon fields. The synchrotron emission and the IC produce very distinguishable bumps in the spectral energy distribution (SED), while the EIC and SSC components cannot always be distinguished.

3C\,279 is a prototypical FSRQ at redshift $z=0.536$, well known for its large-amplitude, rapid variability across the entire electromagnetic spectrum, from radio to Very High Energy (VHE; $E$ > 100 GeV) $\gamma$-rays, and for dramatic polarization-angle swings observed during some flaring episodes \citep[e.g.][]{kiehlmann2016}. Owing to its exceptional activity, 3C\,279 has been one of the best-monitored blazars by {\it Fermi}-LAT and through numerous multi-wavelength (MWL) campaigns, revealing variability on timescales as short as minutes.

3C\,279 was the first FSRQ detected in the VHE $\gamma$-ray band. \citet{magic2008} first reported its detection during a bright state in 2006. 
This discovery triggered extensive modeling efforts aimed at explaining its transparency to VHE photons and the required jet conditions \citep[e.g.][]{sah2012}. The historically brightest $\gamma$-ray outburst in 2015 showed minute-scale flux doubling ($\lesssim$ 5 min), implying very compact emission regions and large Doppler factors, placing stringent constraints on jet magnetization and emission-site geometry \citep{ackermann2016,hess2019,dmytriiev2024}. 

A long-term variability study based on a ten-year MWL dataset \citep{dmytriiev2023} supports the leptonic inverse-Compton interpretation, identifying BLR photons as the dominant target for upscattering. In this study, the pronounced long-term changes in Compton dominance appear primarily driven by variations in magnetic field strength rather than by changes in external photon-field luminosity. While an overall correlation between optical and $\gamma$-ray fluxes is present, 3C\,279 also exhibits ``orphan'' $\gamma$-ray flares, i.e.\ episodes lacking optical counterparts, that remain challenging to reconcile within simple one-zone frameworks.

Between December 2013 and April 2014, the source entered an exceptionally active phase, monitored through a coordinated campaign involving {\it Fermi}-LAT, NuSTAR, Swift, and various ground-based facilities. During this period, multiple distinct $\gamma$-ray flares were detected, exhibiting unusually hard spectra, with photon indices as low as $\Gamma_{\gamma} = 1.7 \pm 0.1$, and flux-doubling timescales of only $\sim$2 hours \citep{Hayashida_2015}. Some flares showed no simultaneous optical variability, indicating an extreme Compton dominance ($L_{\gamma}/L_{\rm syn} > 300$). Modeling these events highlighted serious difficulties for standard one-zone leptonic scenarios, which required very low magnetization, very hard electron distributions, and a large jet power.

On 20 December 2013 (MJD 56646) the FSRQ 3C\,279 was subject to a particular type of flare: a short ($\approx 12$ hours) and strong flare was recorded \citep{Hayashida_2015} in the high-energy $\gamma$-ray band ($20\ \mathrm{MeV} - 300\ \mathrm{GeV}$), but not in the optical and infrared bands. During the flare period, the source was not observed in the X-ray band, leaving us without information in this energy range.

In most studies, flaring episodes are modeled through step-like changes in physical parameters or by invoking transient acceleration episodes. In contrast, here we explore an alternative scenario in which variability arises from changes in the relative geometrical configuration of the emitting region and the external photon fields.

We describe the apparent orphan flare of 20 December 2013 using the time-dependent \texttt{EMBLEM} code \citep{dmytriiev2021}, with an approach that differs from previous scenarii \citep{Hayashida_2015,Lewis2019,Paliya2016} in that it does not involve additional acceleration or time-varying injection of the plasma electrons within the jet, but rather results from the bulk motion and geometry of the external photon fields.

In Section~\ref{sec:observations}, we summarize the available multiwavelength data. The setup of the model and its application to the flare data  is presented in Section~\ref{sec:model}. The results and their implications are discussed in Sections~\ref{sec:results} and~\ref{sec:discussion}.

\section{Multiwavelength observations of the 3C\,279 orphan flare}
\label{sec:observations}
Data available over three days before the flare show 3C\,279 in a quiescent state. We refer to this state as Period LS (MJD 56643--56645). The flare period is referred to as Period F (MJD 56646).

During Period LS, 3C\,279 was observed in $\gamma$-rays by \textit{Fermi}-LAT, in X-rays by \textit{NuSTAR} and \textit{Swift}-XRT, in UV and optical by \textit{Swift}-UVOT, in optical and near-infrared by the Kanata Telescope and SMARTS, and in radio by the Submillimeter Array (SMA). Fewer telescopes observed 3C\,279 during Period F because of the short duration of the flare; for this period only data from {\it Fermi}-LAT, the Kanata telescope and SMARTS are available. Unfortunately, there are no simultaneous X-ray data during the flare.

\begin{figure}
\resizebox{\hsize}{!}{\includegraphics[clip=true]{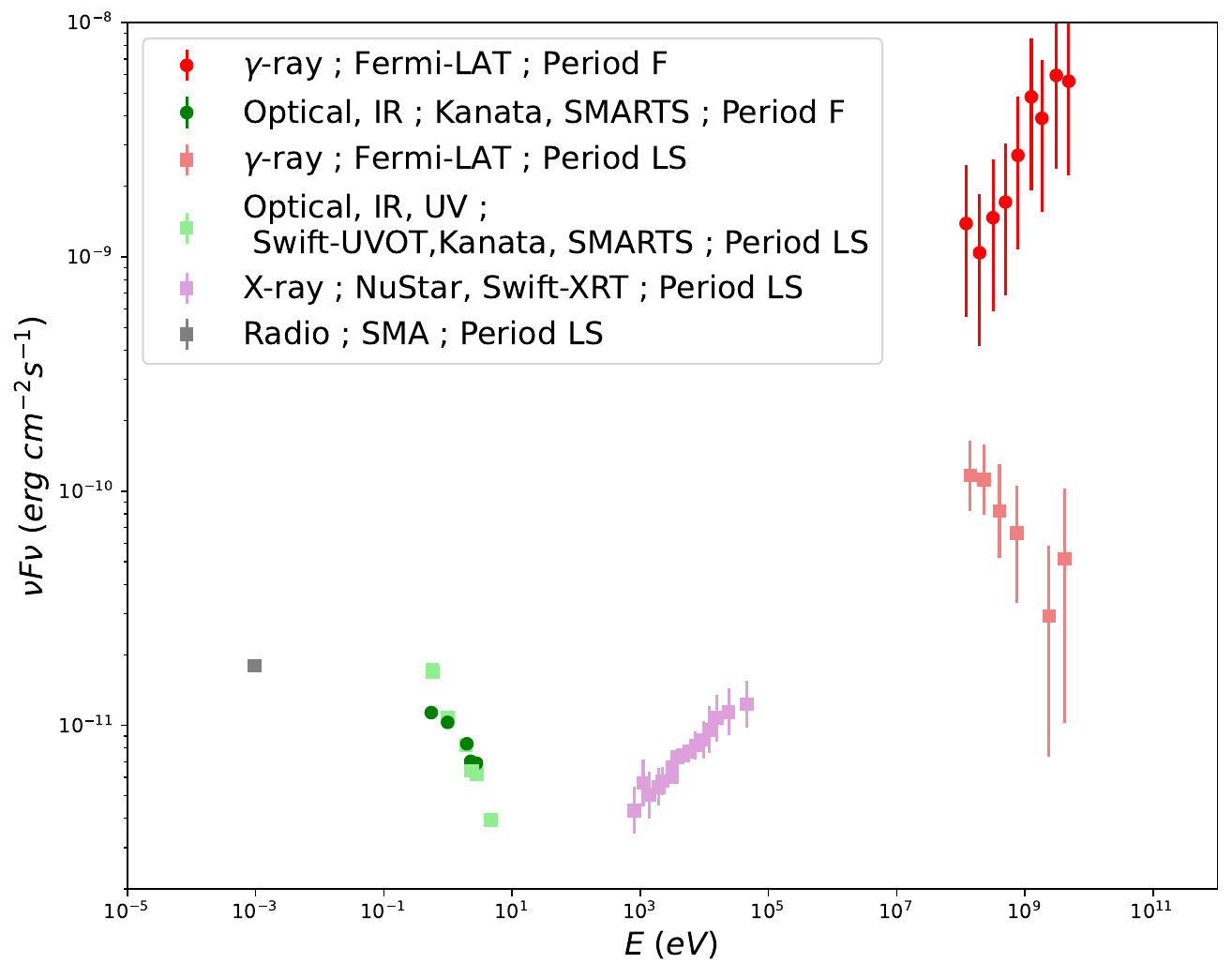}}
\caption{Broadband spectral energy distribution of 3C\,279 for Period LS (low-state) and Period F (flaring state). Error bars are 1$\sigma$ statistical errors. Data points are from \citet{Hayashida_2015}.}
\label{fig:seddata}
\end{figure}

\begin{figure}
\resizebox{\hsize}{!}{\includegraphics[clip=true]{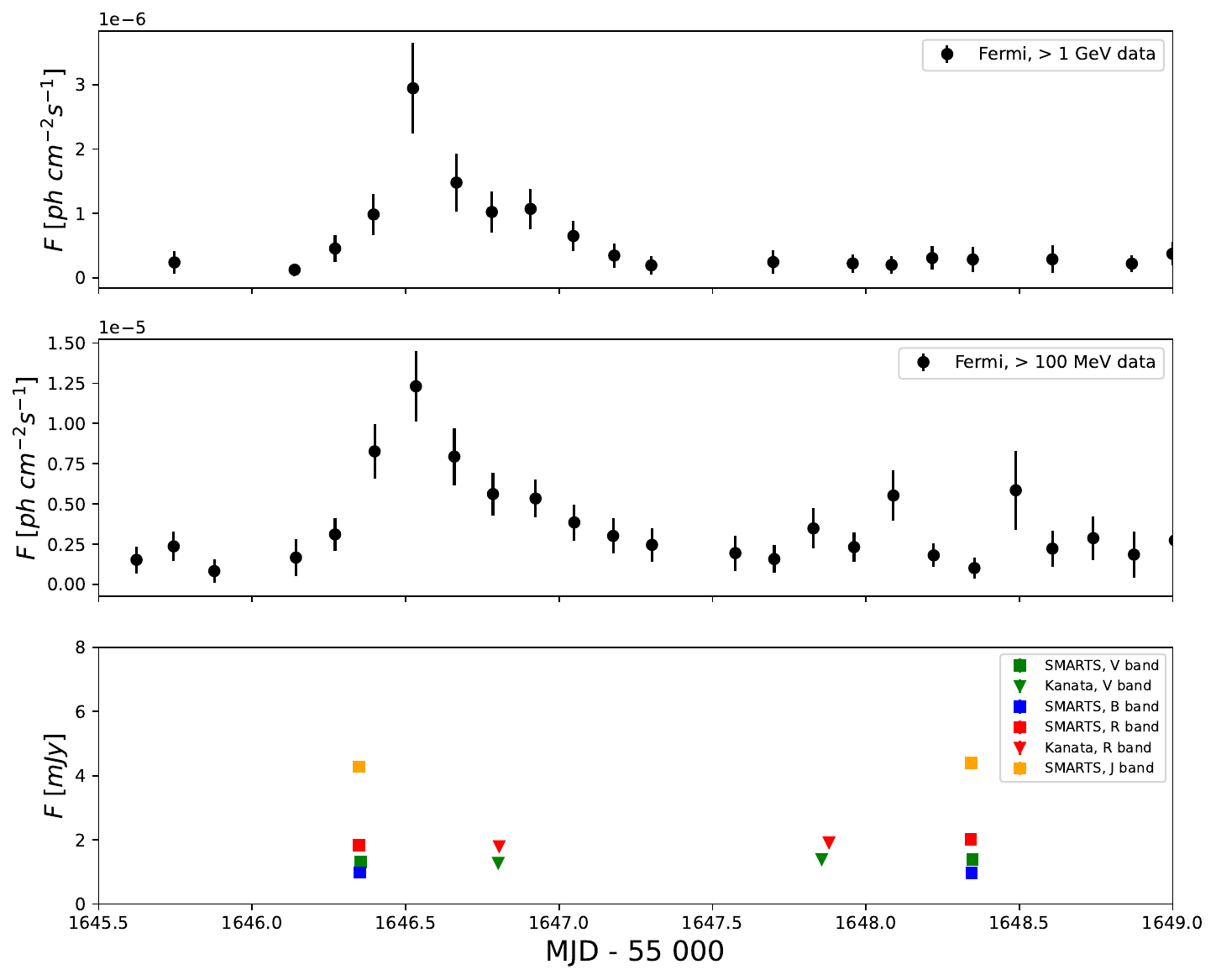}}
\caption{$\gamma$-ray, optical, and infrared light curves of 3C\,279. Top and middle panels show the integrated photon flux for energies $>100\ \mathrm{MeV}$ and $>1\ \mathrm{GeV}$ (192 min bins). Bottom panel displays effective flux density in optical/IR bands. Error bars are 1$\sigma$. Data points from \citet{Hayashida_2015}.}
\label{fig:lcdata}
\end{figure}

The SED and light curve data are taken from \citet{Hayashida_2015} and are plotted in Figs.~\ref{fig:seddata} and \ref{fig:lcdata}. These data exhibit a significant flux variation in the $\gamma$-ray band but not in the optical and infrared bands. The flux variation is described by a large flux increase with a doubling time scale of a few hours, a very hard $\gamma$-ray spectrum and a temporal asymmetry of the light curve with a relatively short rise and slow decay. 

\section{The accelerating blob scenario}
\label{sec:model}

\subsection{Acceleration within the jet}

\citet{acceleration_obs} have recently reported  the presence of a jet with two distinct geometric regimes in the FSRQ\,1928+738: an inner parabolic section transitioning into an outer conical section. They also measured an acceleration of the jet flow within the parabolic region, up to a bulk Lorentz factor $\Gamma \sim 10$, followed by a deceleration further downstream. Since parabolic collimation is theoretically associated with gradual conversion of Poynting flux into kinetic energy, this constitutes the first observational evidence of an acceleration-and-collimation zone (ACZ) in an FSRQ. Such regions have long been predicted by analytical arguments \citep{1987A&A...184..341C,1992ApJ...394..459L,1994ApJ...426..269B,10.1111,Lyubarsky_2009,komissarov_theory,2013ApJ...775..118N} and confirmed in numerical simulations \citep{2008MNRAS.388..551T,2008ApJ...679..990Z,Tchekhovskoy_2009,komissarov_num} and observations of other types of AGNs \citep{2012ApJ...745L..28A,acceleration_obs,2025ApJ...991...75Y,2022A&A...658A.119B,2013ApJ...775..118N,2021ApJ...909...76P,2014ApJ...781L...2A,Hada_2018}. The acceleration is attributed to differential collimation of the jet, wherein different layers follow distinct parabolic profiles, creating a magnetic nozzle. In contrast, within the conical region, the jet undergoes free expansion: all layers scale linearly with distance from the black hole, differential collimation ceases, and acceleration is no longer observed.

Let $r$ denote the jet cross-sectional radius and $z$ the distance from the base of the jet. 
The transition between the parabolic and conical regimes occurs at $z \simeq R_A$, where $R_A$ characterizes the extent of the ACZ. As discussed above, observations and numerical simulations indicate the geometric scaling of the jet envelope:
\begin{equation}
r \propto
\begin{cases}
 z^{1/2}, & z \le R_A, \\
 z, & z \ge R_A,
\end{cases}
\label{radius_relation}
\end{equation}
together with a scaling of the bulk Lorentz factor in the ACZ,
\begin{equation}
\Gamma \propto r \propto z^{1/2}, \qquad (z \le R_A),
\end{equation}
while $\Gamma$ becomes constant or decreases beyond $R_A$.

\subsection{Flaring and low-state emission regions}

To describe the observed orphan flare of 3C\,279, a two-zone model is required. The first zone, hereafter "blob 1", is stationary and is located outside the dusty torus shell, in the jet, and accounts for the quiescent state (Period LS). The second zone, hereafter "blob 2", responsible for the flare emission (Period F), is accelerated from the jet base up to the BLR shell. 

While blob 2 could be interpreted as an overdensity in the plasma injected at the base of the jet, blob 1 might correspond to a standing re-collimation shock or a slowly moving jet component.

\begin{figure}
    \centering
    \begin{tikzpicture}[scale=0.9, line cap=round, line join=round, >=stealth]

    \definecolor{jetblue}{RGB}{130,230,255}
    \definecolor{dustyred}{RGB}{255,120,120}
    \definecolor{blrgreen}{RGB}{50,150,70}
    \definecolor{bloborange}{RGB}{255,150,50}
    \definecolor{blobstat}{RGB}{255,150,150}
    \definecolor{diskyellow}{RGB}{250,220,80}

    \fill[dustyred,opacity=0.35,rounded corners=5pt] (5.5,-1.5) rectangle (7.5,1.5);

    \fill[blrgreen,opacity=0.25,rounded corners=5pt] (2.5,-1.5) rectangle (3.3,1.5);

    \fill[diskyellow,opacity=0.9,rounded corners=3pt] (-0.15,-1.5) rectangle (0.15,1.5);

    \fill[black] (0,0) circle (0.12);

    \path[fill=jetblue,opacity=0.55,domain=0.2:3.0,smooth,variable=\x]
        plot ({\x},{0.25*sqrt(\x)+0.1}) --
        plot[domain=3.0:0.2] ({\x},{-0.25*sqrt(\x)-0.1}) -- cycle;

    \path[fill=jetblue,opacity=0.55]
        (3.0, {0.25*sqrt(3.0)+0.1}) --
        (9.7, {0.25*sqrt(3.0)+0.1 + 0.1*(10-2.5)}) --
        (9.7, {-0.25*sqrt(3.0)-0.1 - 0.1*(10-2.5)}) --
        (3.0, {-0.25*sqrt(3.0)-0.1}) -- cycle;

    \foreach \x/\r/\s in {1.0/0.1/0.3, 2.0/0.141/0.5, 3.0/0.2115/0.8, 4.5/0.317/0.8} {
        \shade[ball color=bloborange,opacity=0.9] (\x,0) circle (\r);
        \draw[->,thick,black] (\x,0) -- ++(\s,0);
    }
    \node[font=\small, black] at (4.6, -0.5) {Blob 2};

    \shade[ball color=blobstat,opacity=0.9] (9,0) circle (0.5);
    \node[font=\small, black] at (9, -0.7) {Blob 1};

    \end{tikzpicture}
    \caption{Schematic view (not to scale) of our scenario. In black the black hole, in yellow the accretion disk, in green the broad line region, in red the dusty torus and in blue the jet. The stationary blob is represented by the pink circle and the moving blob is represented by the orange circles, with arrows showing its velocity. The inner jet is parabolic and the outer jet is conical.}
    \label{fig:AGN_diagram}
\end{figure}
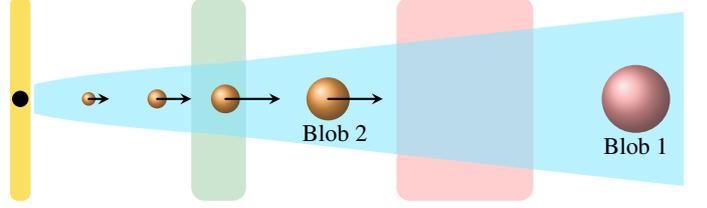

As blob 2 is propagating along the jet, its bulk Lorentz factor is assumed to follow the same acceleration law as the jet flow \citep{ghisellini2009}, 
\begin{equation}
\Gamma_{\mathrm{blob}} = \min\!\left(\Gamma_{\max}, \left(\frac{R_{b-BH}}{3R_S}\right)^{1/2}\right),
\label{acc_blob}
\end{equation}
where $R_{b-BH}$ is the distance of the blob from the black hole, $R_S$ the Schwarzschild radius of the central black hole, and $\Gamma_{\max}$ the terminal Lorentz factor of the jet.

The extent of the ACZ is then given by
\begin{equation}
R_A \equiv 3R_S \Gamma_{\max}^2,
\label{acc_zone}
\end{equation}
meaning that the blob accelerates up to $R_A$, where the jet becomes conical and the blob reaches its terminal velocity. Beyond this point, it propagates with approximately constant $\Gamma_{\max}$.

For both emission regions, the synchrotron, SSC and EIC emissions are calculated. The $\gamma \gamma$- absorption on the Extragalactic Background Light is taken into account according to \citet{dominguez2011}. Internal absorption on the photons reprocessed by the BLR is accounted for, as described in \ref{appendixA}. An upgraded version of the \texttt{EMBLEM} code \citep{dmytriiev2021,dmytriiev2024,2026arXiv260205601T} is used for this application.

\subsection{External photon fields}

The accretion disk, BLR and dusty torus each provide external photon fields that serve as targets for EIC emission. 
The photon field of the accretion disk is modeled following \citet{ghisellini2009} as a multi–temperature black body with a temperature that is a function of radius:
\begin{equation}
    T^4(R) = 
    \frac{3 R_S L_d}{16 \pi \eta \sigma_{\mathrm{MB}} R_{b\text{-}\mathrm{BH}}^3}
    \left[1 - \left(\frac{R}{3R_S}\right)^{1/2}\right],
    \label{eq:Tdisk}
\end{equation}
where $L_d$ is the disk luminosity, $\eta=0.1$ the assumed accretion efficiency, and $R$ the distance from the black hole.  

The disk is modelled as concentric annuli of temperature $T(R)$ and width $\mathrm{d}R$.  
The angle between the blob and a disk annulus is 
\(\theta = \cos^{-1}\!\mu = \cos^{-1}[(1 + R^2 / R^2_{b\text{-}\mathrm{BH}})^{-1/2}]\).  
The disk extends from $R_{\mathrm{in}} = 3R_S$ to $R_{\mathrm{out}}$, corresponding to 
\(\mu_{\mathrm{in,out}} = (1 + R^2_{\mathrm{in,out}} / R^2_{b\text{-}\mathrm{BH}})^{-1/2}\).

The specific radiation energy density in the blob frame is (primes denote the blob frame):
\begin{equation}
    U'_{\mathrm{disk}}(\epsilon', R_{b\text{-}\mathrm{BH}}) =
    \frac{2\pi}{c}
    \int_{\mu_{\mathrm{out}}}^{\mu_{\mathrm{in}}}
    \frac{I(\epsilon' \delta / h)}{\delta}\, \mathrm{d}\mu,
    \label{rad_AD}
\end{equation}
with $\delta = 1 / [ \Gamma_{\mathrm{blob}} (1 - \beta_{\mathrm{blob}} \mu) ]$ the Doppler factor of the blob, $I(\nu)$ the specific intensity of a black body of temperature $T(R)$ and $\epsilon$ the photon energy.

The BLR emits a spectrum of lines that are a result of the reprocessing of a fraction $f_{\mathrm{BLR}}$ of the disk radiation. We model its specific luminosity in the blob reference frame with a sum of Gaussians:
\begin{equation}
    L_{BLR}'(\epsilon') =f_{\mathrm{BLR}}L_d \sum_{i \in \mathcal{L}}  f_i  \frac{1}{\Delta E_i' \sqrt{2 \pi}} \exp\left[-\left(\frac{\epsilon'-E_i'}{\sqrt{2}\Delta E_i'}\right)^2\right],
\end{equation}
with $\mathcal{L} = \{\mathrm{Ly\alpha}, \mathrm{H\alpha}, \mathrm{C\,\textsc{iv}}, \mathrm{C\,\textsc{iii]}},\mathrm{Mg\,\textsc{ii}}\}$ the most significant lines, $f_i$ the fraction of reprocessed luminosity of each line, $E'_i$ the energy of each line in the blob frame and $\Delta E'_i$ the width of each line due to Doppler broadening, in the blob frame. $f_i$ and $\Delta E'_i$ are adapted from \citet{Finke_2016}.

The specific radiation energy density of the BLR is modelled as a spherical shell with a density profile constant within the inner radius of the BLR, $R_{\mathrm{BLR}} = 0.1\sqrt{L_{d,46}} \ \mathrm{pc}$, and decreasing afterwards. In the blob frame it is then \citep{sikora_constraining_2009,hayashida2012}:
\begin{equation}
U'_{\mathrm{BLR}}(\epsilon',R_{b-BH}) = \frac{L'_{\mathrm{BLR}}(\epsilon')\Gamma_{\mathrm{blob}}^2}{3\pi R_{\mathrm{BLR}}^2 c \left[1+\left(\frac{R_{b-BH}}{R_{\mathrm{BLR}}}\right)^{\beta_{\mathrm{BLR}}}\right]},
\label{rad_BLR}
\end{equation}
with $\beta_{\mathrm{BLR}}$ the power law index of the photon energy density of the BLR. As can be seen directly from the above equation, any change in the bulk Lorentz factor of the blob will have an important ipact on $U'_{\mathrm{BLR}}$.

The dusty torus is re-emitting a fraction $f_{\mathrm{DT}}$ of the disk radiation in the form of a thermal distribution of the form of a black body. Thus its specific luminosity in the blob frame is \citep{dermer_equipartition_2014}:
\begin{equation}
    L'_{\mathrm{DT}}(\epsilon') = \frac{15 f_{\mathrm{DT}} L_d}{\pi^4 \epsilon'} \frac{(\epsilon'/\Theta')^4}{\exp(\epsilon'/\Theta')-1}
\end{equation}
with $\Theta' = \Gamma_{\mathrm{blob}} k_{\mathrm{B}}T_{\mathrm{DT}}/m_e c^2$ the reduced temperature of the dusty torus in the blob frame.

The specific radiation energy density of the torus is modelled in the same way as for the BLR and is thus given in the blob frame by \citep{hayashida2012}:
\begin{equation}
U'_{\mathrm{DT}}(\epsilon',R_{b-BH}) = \frac{L'_{\mathrm{DT}}(\epsilon')\Gamma_{\mathrm{blob}}^2}{3\pi R_{\mathrm{DT}}^2 c \left[1+\left(\frac{R_{b-BH}}{R_{\mathrm{DT}}}\right)^{\beta_{\mathrm{DT}}}\right]},
\label{rad_DT}
\end{equation}
with $L_{\mathrm{DT}} = f_{\mathrm{DT}} L_d$, $\beta_{\mathrm{DT}}$ the power law index of the photon energy density of the dusty torus and $R_{\mathrm{DT}} = 2.5\sqrt{L_{d,46}} \ \mathrm{pc}$ the radius of the dusty torus shell.

The X-ray corona is located below and above the disk and re-emits a fraction $f_X \sim 0.3$ of the disk radiation, in the UV and X-ray bands. Its spectrum is assumed to be a cut-off power law: $L_X'(\epsilon') \propto \epsilon'^{-\alpha_X} \exp{(-\epsilon'/\epsilon'_c)}$ (with typical values: $\alpha_X \sim 0$ and $\epsilon_c \sim 100 \mathrm{keV}$). Its radiation energy density in the blob frame is taken from \citet{ghisellini2009}:
\begin{equation}
U'_{\mathrm{X}}(\epsilon',R_{b-BH}) = \frac{L'_{\mathrm{X}}(\epsilon')\Gamma_{\mathrm{blob}}^2}{\pi R_{\mathrm{X}}^2 c} \left(1 - \mu_{\mathrm{X}} - \beta(1 - \mu_{\mathrm{X}}^2) + \frac{\beta^2}{3}(1 - \mu_{\mathrm{X}}^3)\right),
\label{rad_X}
\end{equation}
with $R_{\mathrm{X}}$ the extent of the X-ray coronae, $\beta$ the reduced velocity of the blob and $\mu_{\mathrm{X}} = (1 + R_{\mathrm{X}}^2 / R_{b\text{-}\mathrm{BH}}^2)^{-1/2}$.

\begin{figure}
\resizebox{\hsize}{!}{\includegraphics[clip=true]{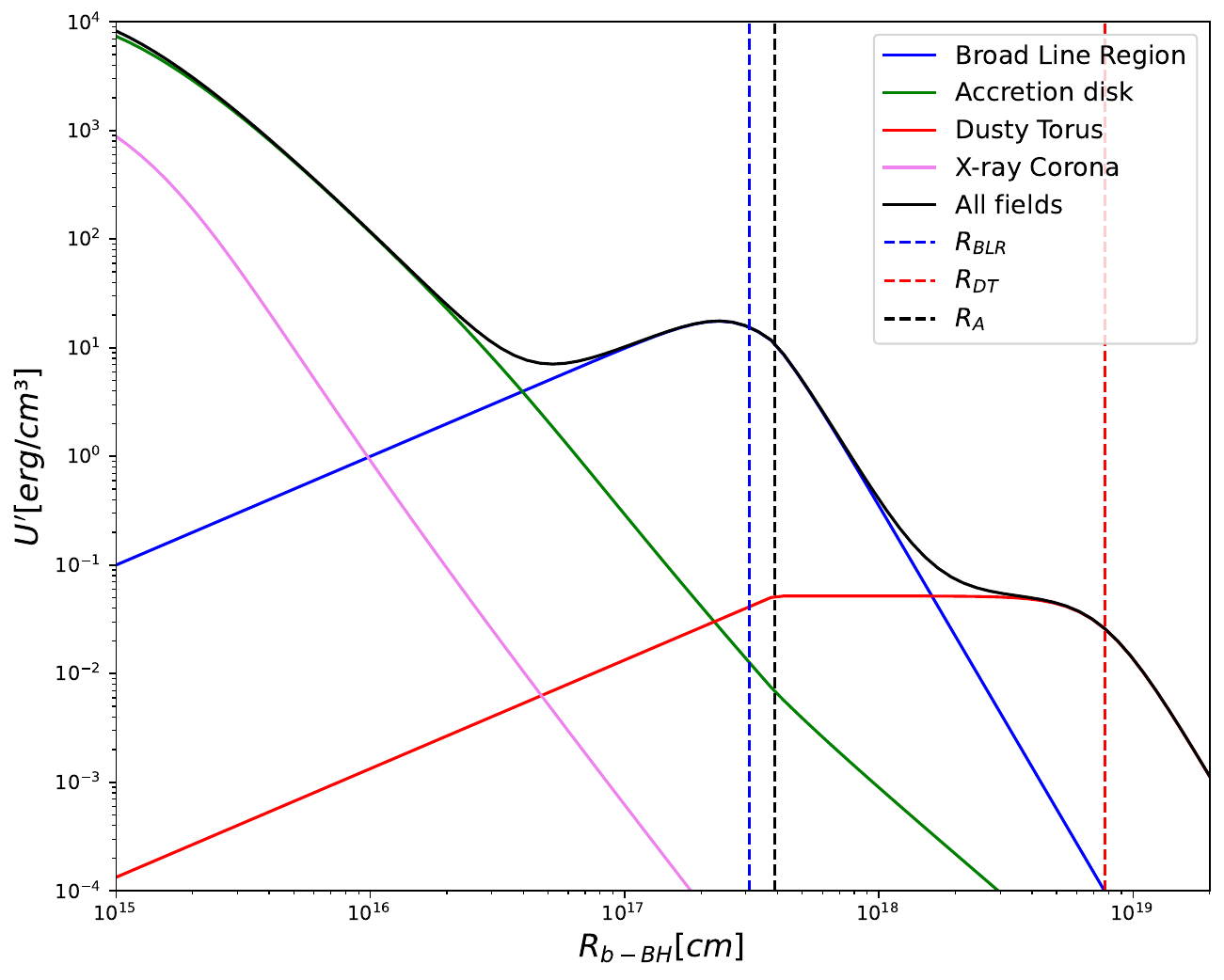}}
\caption{Integrated radiation energy densities of the disk, BLR, and dusty torus in the blob frame as a function of the distance between the blob and the black hole for an accelerating blob. The total radiation energy density is also shown. The radius of the BLR and dusty torus are indicated by vertical dashed lines.}
\label{fig:Uext}
\end{figure}

The different contributions to the radiation energy density (Eqs. \ref{rad_AD}, \ref{rad_BLR}, \ref{rad_DT}, \ref{rad_X}) integrated over energy are plotted in Fig.~\ref{fig:Uext} as a function of the distance between the blob and the black hole for a blob accelerating up until the end of the ACZ zone $R_A$ (Eqs. \ref{acc_blob}, \ref{acc_zone}). The corresponding parameters are listed in Table \ref{t2}.
The radiation energy density originating from the X-ray corona has a negligible contribution compared to the other external photon fields at every distance from the black hole. Thus it is not included in the simulations.
We see that the AD radiation field dominates close to the black hole, the BLR dominates at intermediate distances, and the dusty torus dominates at large distances.
Overall, the total external radiation energy density decreases with distance from the black hole except close to the inner radius of the BLR, where it increases due to the acceleration of the blob,
and close to the inner radius of the dusty torus, where it has a plateau.

\subsection{Evolution of electron populations within the blobs}
The emission is assumed to arise entirely from a population of relativistic electrons (and/or electron–positron pairs) contained within the two plasma blobs. The temporal evolution of this particle population is described by a Fokker-Planck equation. Specifically, the electron spectrum $N_e(\gamma,t) = dN_e/dVd\gamma$ (with $N_e$ the number of electrons) evolves as (\citealp{kardashev1962,tramacere2011}):
\begin{equation}
\begin{split}
\frac{\partial N_e(\gamma,t)}{\partial t} = 
\frac{\partial}{\partial \gamma}\left[\left(\frac{1}{t_{\mathrm{cool}}(\gamma,t)}+\frac{1}{t_{\mathrm{ad}}}\right)\gamma N_e(\gamma,t)\right]\\
- N_e(\gamma,t)\left(\frac{1}{t_{\mathrm{esc}}(t)}+\frac{3}{t_{\mathrm{ad}}}\right)
+ Q_{\mathrm{inj}}(\gamma,t),
\end{split}
\label{FP_eq}
\end{equation}
with $\gamma$ the Lorentz factor of the electrons, $t_{cool}$ the characteristic cooling timescale due to synchrotron and IC emission \citep{dmytriiev2021}, $t_{ad}$ the adiabatic cooling timescale due ti the expansion of the blob \citep{tramacere2022}, and $t_{esc}$ the espace timescale of electrons leaving the blob.
The description of the adiabatic expansion has been added to the original \texttt{EMBLEM} code as detailed in~\ref{appendixB}. 

The injection rate $Q_{\mathrm{inj}}$ is given in the form of a power-law:
\begin{equation}
Q_{\mathrm{inj}}(\gamma,t) = N\left(\frac{\gamma}{\gamma_{\mathrm{pvt}}}\right)^{\alpha_1},
\end{equation}
or in the form of a broken power-law:
\begin{equation}
    Q_{\mathrm{inj}}(\gamma,t) = N\left(\frac{\gamma}{\gamma_{\mathrm{pvt}}}\right)^{\alpha_1} \left(1 + \left(\frac{\gamma}{\gamma_{\mathrm{brk}}}\right)^2\right)^{\frac{\alpha_2-\alpha_1}{2}},
\end{equation}
for $\gamma_{\min}\le\gamma\le\gamma_{\max}$ and accounts for the injection of electrons in the blob with the shape of a (broken) power law spectrum of norm $N$ and indexe(s) $\alpha_1(, \alpha_2) < 0$.

We assume that relativistic electrons are continuously injected at a constant rate in both emission regions, without specifying the underlying mechanism. For blob 1, a standing shock might cause the injection, while for the moving blob 2 the source of relativistic electrons might be found in turbulent or shear acceleration, or in a acceleration on a bow shock, as the plasma blob moves along the jet.

\subsection{A flare from a blob crossing the BLR}

\begin{figure}
\resizebox{\hsize}{!}{\includegraphics[clip=true]{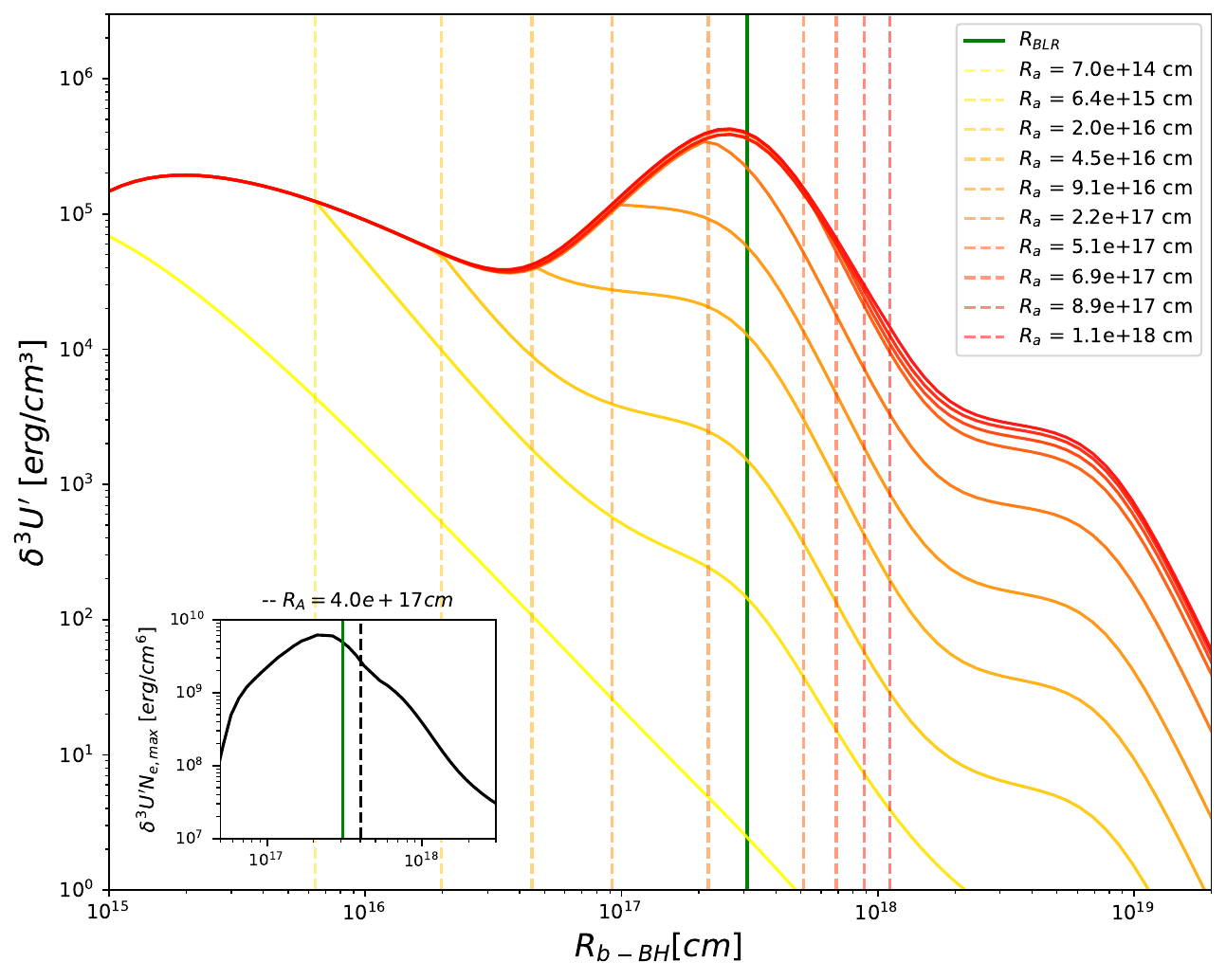}}
\caption{Total radiation energy density (disk + BLR + dusty torus), boosted by $\delta^3$.
Both the radiation field and $\delta$ evolve with the distance from the black hole $R_{b\text{-}BH}$. The different colors indicate different sizes of the acceleration zone (the dashed lines of corresponding colors mark the extent of each acceleration region). The inset is the total radiation energy density boosted by $\delta^3$ and multiplied by the maximum of the electron distribution as computed in Section \ref{sec:results}, for an acceleration radius $R_A = 4.0 \times 10^{17} \, \mathrm{cm}$.}
\label{fig:delta3U}
\end{figure}

As blob 2 moves along the jet, the time-dependent total external radiation field in its rest frame is given by
\begin{equation}
U'_{\mathrm{tot}} = U'_{\mathrm{disk}} + U'_{\mathrm{BLR}} + U'_{\mathrm{DT}},
\end{equation}
 Since the EIC flux depends linearly on $U'_{\mathrm{tot}}$, any variation in the external radiation field directly affects the observed emission. At first glance, as shown in Fig.~\ref{fig:Uext}, one would expect the flux to decrease as the blob moves farther from the black hole. However, this neglects the fact that the observed flux is Doppler-boosted by a factor $\delta^3$ 
 , which evolves with the variation of the bulk Lorentz factor according to Eq.~\ref{acc_blob}. Therefore, the relevant quantity for the observed EIC flux is $\delta^3 U'_{\mathrm{tot}}$, which is plotted in Fig.~\ref{fig:delta3U}. In this plot, $\delta^3 U'_{\mathrm{tot}}$ is computed using Eqs.~\ref{acc_blob} and \ref{rad_AD}--\ref{rad_DT}, for different sizes of the acceleration region in the jet, parameterized by the radius $R_A = 3 R_S \Gamma_{\mathrm{max}}^2$.

The behaviour of $\delta^3 U'_{\mathrm{tot}}$ depends sensitively on the size of the jet acceleration region, $R_A$.
If the acceleration region is compact ($R_A \ll R_{\mathrm{BLR}}$), $\delta^3 U'_{\mathrm{tot}}$ decreases monotonically with distance, and thus only a decay of the EIC flux is expected. If the acceleration region is comparable to the BLR size ($R_A \sim R_{\mathrm{BLR}}$), $\delta^3 U'_{\mathrm{tot}}$ reaches a maximum near $R_A$. Finally, if the acceleration region extends beyond the BLR ($R_A \gtrsim R_{\mathrm{BLR}}$), the maximum of $\delta^3 U'_{\mathrm{tot}}$ occurs near the BLR radius, $R_{\mathrm{BLR}}$. In these latter two cases, a flare in the EIC flux is expected as the blob crosses the region where $\delta^3 U'_{\mathrm{tot}}$ peaks, i.e.\ before or while crossing the BLR.

The electron cooling timescale is also evolving during the propagation of the blob along the jet. As the blob approaches the broad line region, the external radiation fields become stronger in its rest frame, leading to a shorter cooling timescale. In this regime, the electrons lose energy efficiently, resulting in a lower internal electron energy density $u'_e$. Once the blob crosses the BLR, the external radiation field rapidly weakens, causing the electron cooling timescale to increase. As a consequence, $u'_e$ gradually builds up again, leading to an enhancement of both the synchrotron and SSC emissions. A delayed flare is therefore expected in these components, occurring after the blob exits the BLR, unless it is slowed down. Because the blob simultaneously undergoes adiabatic expansion while propagating along the jet, this increase in emission should eventually be followed by a gradual decay. The impact on the evolution of the electron distribution is illustrated in more detail in~\ref{appendixC}.

One might wonder whether the evolution of the electron distribution could alter the 
qualitative behaviour of the EIC flux with respect to the $\delta^3 U'_{\rm tot}$ proxy. In the 
cooling-dominated regime, one analytically expects $N_{e} \propto 1/U'_{\rm tot}$ 
\citep{dmytriiev2021}, assuming EIC losses dominate, so that $\delta^3 U'_{\rm tot} N_{e} 
\propto \delta^3$, which increases as the blob approaches the BLR. Far beyond the BLR, where 
cooling becomes subdominant with respect to adiabatic expansion and escape, $N_{e}$ 
decreases together with $U'_{\rm tot}$, and thus $\delta^3 U'_{\rm tot} N_{e}$ also decreases. The 
qualitative picture, namely an initial rise followed by a decay, is therefore preserved.

However, the behaviour near the BLR crossing, where the transition between these two regimes occurs, 
cannot be resolved analytically due to the complexity of the full kinetic equation.
To assess its impact, we make use of the numerically computed time-dependent electron distributions and extract the evolution of their peak, $N_{e,\mathrm{max}}$ (see Section \ref{sec:results} for the parameters used and ~\ref{appendixC}), which occurs within the range of Lorentz factors that dominate the production of the observed GeV emission via external Compton scattering on BLR photons. In the inset of Fig.~\ref{fig:delta3U}, we therefore show the evolution of the quantity $\delta^3 U'_{\rm tot} N_{e,\mathrm{max}}$. As shown in ~\ref{appendixC}, $N_{e,\mathrm{max}}$ does not strictly follow a $1/U'_{\rm tot}$ scaling, due to the additional contributions from adiabatic expansion, synchrotron, and SSC cooling, which, although subdominant, cannot be fully neglected. However, the total variation of $N_{e,\mathrm{max}}$ with distance remains modest (a factor of $\sim$3--4), much smaller than the order-of-magnitude variations of $\delta^3U'_{\rm tot}$ over the same range. As a result, $\delta^3 U'_{\rm tot} N_{e,\mathrm{max}}$ displays the same qualitative rise-and-fall behaviour as $\delta^3 U'_{\rm tot}$ alone. This confirms that the latter provides a valid proxy for the EIC flux profile and that the detailed evolution of the electron distribution does not qualitatively alter our conclusions.

To summarize, a flare in the EIC $\gamma$-ray emission is expected to peak, when the blob is located near or just inside the BLR ($R_{\mathrm{BLR}}$), followed by a delayed flare occurring after the blob has crossed the BLR, associated with the rise of the synchrotron and SSC components.

For the observer, the time required for the blob to traverse the acceleration region, corresponding to the typical duration of the observed EIC flare, is:
\begin{equation}
    t_{\mathrm{obs}} = \frac{R_A (1+z)}{\beta_{\mathrm{blob}} c \, \delta \, \Gamma_{\mathrm{blob}}} ,
\end{equation}
We note the presence of both $\Gamma_{\mathrm{blob}}$ and $\delta$ in the denominator, due to a Lorentz contraction between the jet and blob frames, followed by a Doppler shift between the blob and observer frames \citep[e.g.][]{Finke_2016}.
Assuming typical parameters for 3C\,279, namely $\Gamma_{\mathrm{blob}} \simeq \delta \simeq 30$ and $R_A \simeq 3\times10^{17}\,\mathrm{cm}$, we obtain
$t_{\mathrm{obs}} \sim 6~\mathrm{h}$,
which is consistent with the duration of the rapid orphan $\gamma$-ray flare observed in this source on MJD\,56646.

\subsection{Model parameters}

The mass of the central supermassive black hole and the redshift of the source are fixed parameters with values from \citet{massbh} and \citet{redshift}. The disk luminosity is set to $\sim 10\%$ of the Eddington luminosity. The fraction of the radiation re-emitted by the dusty torus is set to $f_\mathrm{DT} = 0.3$ \citep[see ][]{frac_DT} and the fraction for the BLR is fixed to the ad-hoc value $f_\mathrm{BLR}=0.1$. The dusty torus power law index is set to $4$ \citep{hayashida2012}. The BLR power law index depends on the width of the BLR region. In \citet{hayashida2012} it is set to $3$ and in \citet{Finke_2016} it is $7.7$. In our model, we set it to an empirical value of $4$, allowing us to reproduce the slow decay time scale of the $\gamma$-ray light curve after the peak of the flare. 

The angle between the jet and the line of sight $\theta_j$ and the intrinsic opening angle of the conical jet $\theta_o$ are based on measurements by \citet{pushkarev2017}. The maximal jet Lorentz factor, i.e.\ the Lorentz factor in the conical part of the jet, is computed following the procedure in \citet{pushkarev_jet_2009}:  $\Gamma_{\mathrm{max}} \sim 0.26 / \theta_o$.
The other parameters are free and taken to best fit the observed points. However, the size of blob 2 (responsible for the flare) is constrained by the causality condition:
\begin{equation}
    R^{(2)}_b \leq \frac{ct_{\mathrm{var}}\delta}{1+z},
\end{equation}
with the observed variability time scale $t_{\mathrm{var}}$ taken to be a few hours.
We do not attempt to model the very rapid variability of the source detected during certain flare periods, which has to be ascribed to a different mechanism, possibly arising from a more magnetised region in the base of the jet.

Since blob 1 is located in the conical part of the jet, its size is also constrained by $R_b^{(1)} \lesssim R_{b-BH}\theta_o$. The distance between the black hole and the blobs are free parameters but blob 2 must initially be located between the BLR and the black hole for a flare to take place. The result is not sensitive to the precise position of the initial location, since blob 2 does not contribute significantly to the emission during the early phase of the jet acceleration where its bulk Lorentz factor is still small.

One can check that with these parameter values, we have an acceleration zone of size $R_A = 3R_S\Gamma_{\mathrm{max}}^2 \sim 4 \times 10^{17} \mathrm{cm}$ and that the BLR is located at $R_{\mathrm{BLR}} \sim 3.1 \times 10^{17} \mathrm{cm}$, such that a flare is expected.

The size, magnetic field and distance from the black hole of blob 2 evolve because it moves in the jet, following Eq.\ref{acc_blob} and undergoes adiabatic expansion, as detailed in~\ref{appendixB}.

All model parameter values used in the simulations are listed in Tables~\ref{t1} and \ref{t2}.

\begin{table}[h!]
\caption{\label{t1}Parameters used in the simulation of emission from the two blobs. [a,b] means during the simulation, the parameter goes from the initial value a to the value b.}
\centering
\begin{tabular}{lcc}
\hline\hline
Parameter&Blob 1&Blob 2\\
\hline
&Blob characteristics&\\
\hline
Type &Stationary &Accelerating\\
$B(\mathrm{G})$ &0.09&[0.03,0.02]\\
$R_{b-BH} (\mathrm{cm})$& $2.0 \times 10^{19}$ & $[0.05,3] \times 10^{18}$ \\
$R_b (\mathrm{cm})$ &$5.0 \times 10^{16}$&$[2.0,2.9] \times 10^{15}$ \\
\hline
&Injection spectrum& \\
\hline
Type     &BPL&PL\\
$N(\mathrm{cm^{-3}s^{-1}})$&$3.0 \times 10^{-7}$   &$5.0 \times 10^{-1}$\\
$\gamma_{\mathrm{pvt}}$ & $215$ & $200$ \\
$\gamma_{\mathrm{brk}}$ & $3000$ & / \\
$\alpha_{\mathrm{1}}$ & $-2.5$ & / \\
$\alpha_{\mathrm{2}}$ & $-3.5$ & $-3.5$ \\
$\gamma_{\mathrm{min}}$ & $200$ & $1300$ \\  $\gamma_{\mathrm{max}}$ & $10^5$ & $10^5$ \\
\hline
\end{tabular}
\end{table}

\begin{table}
\caption{Fixed AGN parameters used for the model.}
\label{t2}
\centering
\begin{tabular}{lc}
\hline\hline
Parameter & Values \\
\hline
Redshift ($z$) & $0.536$ \\
Maximal Lorentz factor ($\Gamma_{\mathrm{max}}$) & $30$ \\
Opening angle ($\theta_o$) & $0.5^\circ$ \\
Jet angle ($\theta_j$) & $1.91^\circ$ \\
Black hole mass ($M_{BH}$) & $5 \times 10^8\,M_\odot$ \\
Disk luminosity ($L_d$) & $1.0 \times 10^{46}\ \mathrm{erg\ s^{-1}}$ \\
BLR fraction ($f_{\mathrm{BLR}}$) & $0.1$ \\
DT fraction ($f_{\mathrm{DT}}$) & $0.3$ \\
BLR power law index ($\beta_{\mathrm{BLR}}$) & $4$ \\
DT power law index ($\beta_{\mathrm{DT}}$) & $4$ \\
\hline
\end{tabular}
\end{table}

\section{Results}
\label{sec:results}

We compute the evolution of the electron populations in the two blobs (Eq.~\ref{FP_eq}) and the resulting synchrotron, SSC and EIC emissions, while taking into account the evolving external radiation fields (Eqs.~\ref{rad_AD}–\ref{rad_DT}), the evolving bulk Lorentz factor of blob~2 (Eq.~\ref{acc_blob}), the adiabatic expansion of blob~2 
and $\gamma\gamma$ absorption of the emitted spectrum in the BLR.
Radiative transfer between the two blobs is neglected. After conducting the simulations, we verified that, in the FSRQ rest frame, the luminosities of the individual blobs ($L^{1,2} < 10^{40} \mathrm{erg/s}$ at all times) are negligible compared to the accretion-disk luminosity. Consequently, the external photon field contributed by one blob to the other is insignificant, justifying the neglect of radiative transfer between the blobs.

\begin{figure}
\resizebox{\hsize}{!}{\includegraphics[clip=true]{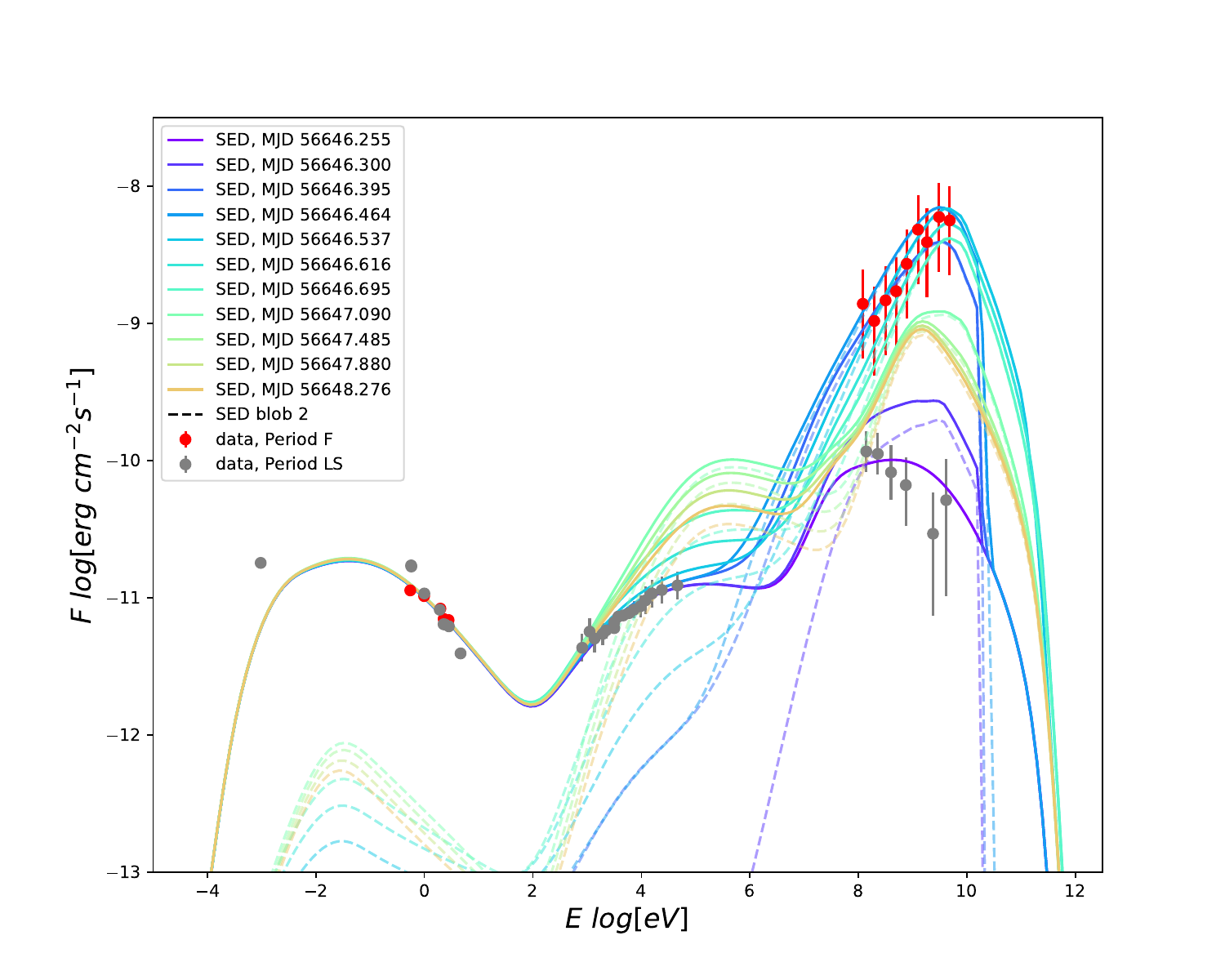}}
\caption{Simulated SED of the total emission of both blobs. The emission of blob~1 completely dominates the Period LS; EIC emission from blob~2 accounts for the high-energy flux points ($E\gtrsim 0.1\ \mathrm{GeV}$) of Period F. The emission of blob 2 is also plotted in dashed lines.}
\label{fig:sedmodel}
\end{figure}

The time evolution of the particle populations and the resulting spectral energy distributions (SEDs) and light curves are computed for each blob and summed up.  Fig.~\ref{fig:sedmodel} shows the combined SED of both emission regions for several time steps. Blob~1 dominates the synchrotron emission in the optical band and the SSC emission in the X-ray band at all times. It reproduces the high-energy bump during the low state. Emission from blob~2 is responsible for the EIC flare at high energies. Above the {\it Fermi}-LAT data points, the effect of $\gamma$-$\gamma$ absorption can be seen in the SED as a sharp cut-off for the contributions from blob~2 before it reaches the inner radius of the BLR shell.

\begin{figure}
\resizebox{\hsize}{!}{\includegraphics[clip=true]{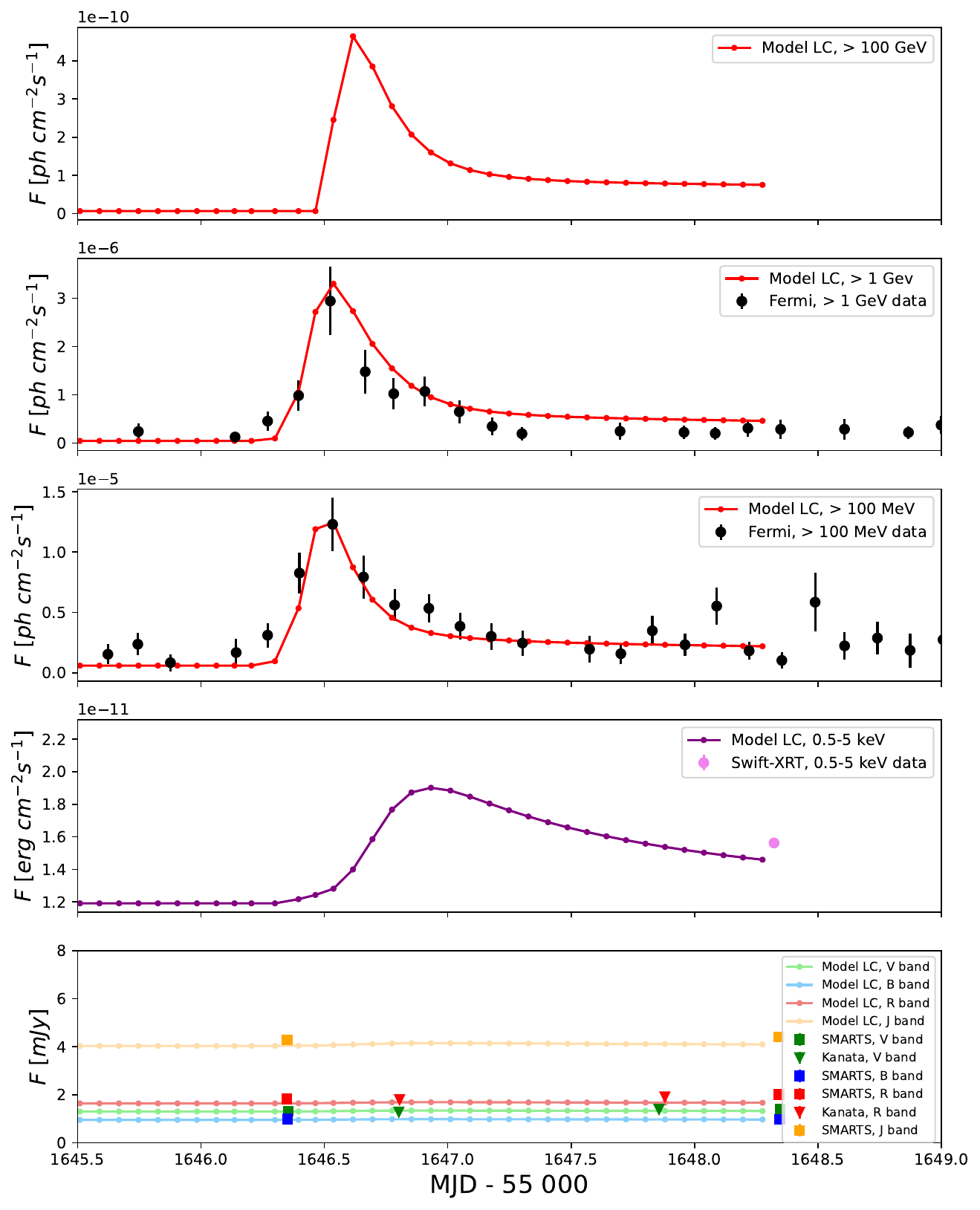}}
\caption{Simulated light curves in different bands ($E>100\ \mathrm{GeV}$, $E > 1 \ \mathrm{GeV}$, $E>100\ \mathrm{MeV}$, $E = 0.5 - 5 \ \mathrm{keV}$, optical bands and infrared J band) from the emissions of both blobs.}
\label{fig:lcmodel}
\end{figure}

Fig.~\ref{fig:lcmodel} presents the simulated multi-band light curves, demonstrating the rapid rise and slower decay of the high-energy flare, while the low-energy bands remain approximately constant. 

The scenario based on one stationary and one accelerating blob yields a satisfactory reproduction of the SED and multi-wavelength light curves. The acceleration of blob~2 up to a distance not far above the inner radius of the BLR leads to a rapid increase of the EIC emission and a steep rise in the high-energy light curves. The subsequent slow decay of the light curve reflects the decrease of the BLR photon density at larger radii. The dominant synchrotron emission from the stationary blob~1 leads to a constant flux at low energies.

The predicted delayed flare from the synchrotron and SSC emission of blob 2 appears in the X-ray band, although it remains hidden at lower frequencies due to the dominant contribution from blob 1. Unfortunately, no X-ray observations are available during the $\gamma$-ray flare itself, preventing a direct comparison with this prediction. The Swift-XRT observation at MJD 56648.3 lies above the declining model light curve, but it was taken nearly two days after the flare and may correspond to unrelated variability or to a distinct flaring episode. Therefore, its relation to the event modeled here cannot be firmly established.

Although no data are available in the very-high-energy band for this flare, a light curve has been added to show the expected sharp and delayed flux variations above $100\ \mathrm{GeV}$, assuming a sufficiently high $\gamma_{max}$ of the electrons in blob~2 to produce emission in this band.

As blob 2 moves beyond the inner BLR shell,  $\gamma\gamma$ absorption drops significantly (cf.~\ref{appendixA}). The associated variability timescale is therefore very short ($\sim 1 \ \mathrm{hour}$), and depends on the width of the BLR.

The total power originating from the two emitting zones can be approximated by:
\begin{equation}
\begin{split}
    L_{\mathrm{tot}}(t) = L_1 + L_2(t) \approx 
    \pi c \left(R_b^{(1)} \Gamma_{\mathrm{blob}}^{(1)}\right)^2 \left[u_e'^{(1)} + u_B'^{(1)}\right] \\
    + \pi c \left(R_b^{(2)}(t)\,\Gamma_{\mathrm{blob}}^{(2)}(t)\right)^2 \left[u_e'^{(2)}(t) + u_B'^{(2)}(t)\right].
\end{split}
\end{equation}

The maximum inferred luminosity occurs when the electron energy density in blob 2 reaches its peak, shortly after the blob has crossed the BLR. We obtain
\[
L_{\mathrm{tot,max}} \simeq 2.1 \times 10^{46}~\mathrm{erg\,s^{-1}} \sim \frac{1}{3} L_{\mathrm{Edd}}.
\]
Although the required luminosity during this exceptional flare is clearly very high, the total energy budget remains well below the Eddington luminosity of the accreting black hole.

\section{Discussion}
\label{sec:discussion}

\subsection{Constraints from VLBI observations of 3C\,279}

The structure of the radio jet of 3C\,279 has been particularly well studied over a range of size scales with VLBI and space-VLBI observations. Even though the small viewing angle prevents a detailed mapping of its deprojected subparsec scale structure, which is the most relevant for the supposed flare emission in our model, several constraints from those observations can be compared with our scenario.

According to \citet{Homan2009} and \citet{Bloom2013}, the analysis of the projected subparsec scale jet at 15\,GHz with the VLBA shows several accelerating knots, which are attributed both to changes in the Lorentz factor and the viewing angle. These movements do not seem to be connected to a single underlying flow speed, but more likely to shocks in the jet or jet bending. A jet closely aligned with the observer and bent by small angles is also proposed to explain the movements of the radio core components observed at 230\,GHz with the EHT~\citep{Kim2020}. This resolved core region would correspond to a deprojected distance along the jet of around 5\,pc, i.e.\ far downstream from our assumed flaring blob 2. Acceleration of a jet component by a propagating shock might be an alternative mechanism to the bulk acceleration through differential collimation we are assuming. Such an alternative could still be covered within our proposed phenomenological scenario.

Observed Lorentz factors for individual components resolved with VLBI range from $\Gamma \sim 10 - 40$ \citep{Bloom2013, Homan2009, Jorstad2017}, while \citet{Lister2016} set the upper limit to $\Gamma \sim 50$. These observations are consistent with the maximum Lorentz factor assumed for the components in our scenario.

The core-separation map compiled with VLBI data from the MOJAVE survey \citep{Lister2018, Lister2021} does not show any clearly identified moving radio knot that could be associated with the flaring event of December 2013, although a large number of radio knots are observed at different core separations up to about 2\,mas
starting from 2014.

Constraints on the extension of the jet acceleration region have become more stringent thanks to EHT data. Based on VLBA observations, it cannot be clearly distinguished whether the jet acceleration is completed on subparsec scales or continues at parsec scales and beyond \citep{Homan2009}. Observations of the nuclear region with the ELT seem to indicate that the intrinsic acceleration region of the jet should be located well upstream, i.e.\ around 0.20\,pc projected distance translating to around 6\,pc along the jet from its base \citep{Kim2020}. As was shown in Fig.~\ref{acc_zone}, the exact extension of the acceleration region in our scenario is not strongly constrained, as long as it is located beyond the BLR radius. A distance of a few parsec is thus compatible with the proposed model.

If the acceleration region extends to several parsec and is marked by a transition from a parabolic to a conical jet, the stationary emission region blob~1, located at a distance of about $6.5$\,pc in our scenario, which is not strongly constrained, could be interpreted as a shock that would be naturally expected at such a transition. 

On the other hand, \citet{Burd2022} find evidence in VLBI data for a transition from a parabolic to a conical jet shape beyond the Bondi sphere in several blazars, including 3C\,279. In the latter, the transition point is estimated to be located at $0.61 \pm 0.22$\,mas, associated with a deprojected distance of 97\,pc for a viewing angle assumed to be 2.4\,deg by the authors. If this change in the jet geometry marks indeed the transition from a magnetically dominated to a particle dominated jet, acceleration would be expected to occur over a much larger distance than derived from the EHT data on the nuclear region. In this case, our scenario may still be valid if one assumes a stratified spine-in-sheath jet structure with separate emission regions for the high-energy radiation and the radio components.

While standard jet propagation models generally assume a magnetically dominated jet up to about $10^5 \, R_s$ \citep{Kim2020}, i.e.\ to a deprojected distance of several \,pc, the two emission regions in our proposed scenario are strongly dominated by the kinetic energy density of the particles
($u'^{(2)}_e / u'^{(2)}_B \sim 10^6$, $u'^{(1)}_e / u'^{(1)}_B \sim 10^2$).
This is a general requirement for the observed strong Compton dominance to emerge.
The very small magnetisation of blob~2 is primarily driven by the X-ray flux upper limit 
around MJD~56648.3. Should particle injection become less efficient after the BLR crossing, 
this constraint would be relaxed, and a magnetisation as high as $u'^{(2)}_B / u'^{(2)}_e 
\sim 10^{-4}$ could be achieved, as verified by an additional simulation ($B^{(2)} = 0.3$~G, 
injection suppressed at $R = 10^{18}$~cm). However, this comes at the cost of an additional 
ad hoc parametrisation, and we therefore retain the original scenario.

While no direct observational constraints are available for the magnetic field strength of 
blob~2, blazar jets are generally assumed to be magnetically dominated close to the base. 
The local magnetic field of the moving emission region is thus not necessarily representative 
of the large-scale jet conditions: localized dissipation processes are expected to strongly 
modify the plasma properties, converting magnetic energy into particle energy and naturally 
producing compact, particle-dominated regions with $u'_e \gg u'_B$. Scenarios based on magnetic reconnection, such as the 
striped-jet mechanism proposed by \citet{Zhang_2021} could provide a physically motivated 
framework in which efficient particle acceleration and reduced local magnetisation arise 
naturally from magnetic reconnection, without implying a globally weakly magnetised flow.

Alternatively, blob~2 could represent an overdense, weakly magnetised plasma structure 
injected into the surrounding magnetised jet, possibly during the accretion process, as already pointed out by \citet{Hayashida_2015}. A clear 
understanding of the physical details would require a thorough exploration through simulations, 
which is beyond the scope of the current work.

Core-shift measurements based on VLBI data from the MOJAVE survey put the location of the radio core at $15.4$\,GHz at a distance of less than $\sim 8$\,pc and its magnetic field strength at $> 0.05$\,G, while the later is limited to be $<0.42$\,G at a distance of $1$\,pc \citep{Pushkarev2012}. The analysis of {\it RadioAstron} data by \citet{Toscano2025} yields a value of $\sim 0.2$ \,G near the $22$\,GHz radio core. These observations are broadly consistent with the assumed magnetic field strength of $0.09$\,G for blob 1 at a distance of several parsec.

\subsection{Comparison with previous interpretations}

The exceptional $\gamma$-ray flare of 3C~279 observed in December 2013 has motivated several theoretical studies, reflecting the difficulty to explain its extreme properties within standard blazar emission scenarios. The combination of a hard $\gamma$-ray spectrum, rapid variability, high Compton dominance, and the absence of contemporaneous optical activity places strong constraints on the physical conditions of the emitting region.

\cite{Paliya2016} investigated this event using both a time-dependent lepto-hadronic model and a two-zone leptonic radiative scenario. In the leptonic interpretation, the observed broadband emission arises from two spatially distinct regions: a compact, fast-moving zone responsible for the $\gamma$-ray emission, and a larger region accounting for the IR-to-X-ray radiation. These regions differ from the single emitting blob that characterizes the quiescent state prior to the event. In the lepto-hadronic framework, the high-energy component is dominated by proton synchrotron radiation, while the low-energy emission originates from electron synchrotron processes. Although both approaches successfully reproduce the observed spectral energy distribution, they require either three emission zones operating quasi-simultaneously or energetically demanding proton populations.

\citet{Hayashida_2015} interpreted this $\gamma$-ray flare within a one-zone leptonic external Compton scenario, placing the emitting region inside the BLR. The preceding quiescent state was modeled with a one-zone leptonic Compton scenario, effectively resulting in an overall two-zone description. The absence of correlated optical variability, together with stringent upper limits on the synchrotron self-Compton component, implies a very high Compton dominance and a low magnetic energy density, indicating a strongly matter-dominated emitting region, consistent with our model. Their modeling further requires an unusually hard injected electron spectrum extending to high Lorentz factors, which poses challenges for standard particle acceleration mechanisms.

\citet{Lewis2019} revisited the same flaring episode using a one-zone leptonic model in which the electron distribution is obtained from a steady-state Fokker--Planck equation including particle acceleration, escape, and radiative and adiabatic losses. The observed spectral evolution is reproduced through changes in the balance between these processes across different epochs and in different regions in the jet. While this approach demonstrates that first- and second-order Fermi acceleration can account for the hard $\gamma$-ray spectrum without invoking magnetic reconnection, it still relies on exceptionally hard electron distributions and a highly compact, weakly magnetized emission region.

In all of these studies, the flaring $\gamma$-ray emission is assumed to originate within or close to the BLR, while the quiescent emission is typically associated with regions farther downstream in the jet where the dusty torus radiation field dominates. A low magnetisation of the flaring region is also a common feature of these models. Despite these shared geometrical and magnetic assumptions, the physical mechanisms invoked to account for the flare in our work differ substantially.

Our model distinguishes itself from these previous works by providing, to our knowledge, the first fully time-dependent interpretation of the flare that explicitly solves for the temporal evolution of the electron energy distribution and the physical conditions in the emission region. It simultaneously accounts for both the bulk acceleration and the spatial displacement of the emitting zone within the jet. By modeling the physical propagation of a plasma blob through the varying external photon fields of the BLR, we provide a dynamical interpretation of the flare’s origin. This approach naturally reproduces the observed asymmetric $\gamma$-ray light curve and predicts a delayed X-ray enhancement as the emitting region exits the BLR, a feature that is not explicitly addressed in previous steady-state or purely stochastic acceleration models.

Beyond the specific application to the 3C\,279 data set, certain radiative models treat the evolution of the blob parameters as it propagates along the jet \citep[e.g.][]{Zacharias2023}, but the acceleration of the emission region is generally not considered.

\subsection{Applicability of our model to other flares}

Although developed here for a specific orphan $\gamma$-ray flare of 3C\,279, the physical scenario explored in this work is expected to be applicable to a broader class of high-energy flares in FSRQs. The key ingredient of the model is the propagation of a compact emitting zone through a jet embedded in spatially stratified external radiation fields, as naturally expected in sources hosting a BLR and a gradually accelerating flow. In such systems, pronounced $\gamma$-ray flares can arise from purely geometrical and radiative effects associated with bulk motion and displacement along the jet, without invoking impulsive changes in particle acceleration or injection. This makes the scenario particularly relevant for $\gamma$-ray–dominated events with weak or absent low-energy counterparts, which are frequently reported in bright FSRQs.

A distinctive observational signature of this class of flares is an asymmetric GeV $\gamma$-ray ({\it Fermi}-LAT) light curve, characterised by a rapid rise followed by a smoother and more prolonged decay.

The degree of asymmetry is, in our scenario, directly linked to the spatial extent of the acceleration region and to the gradient of the external radiation fields along the jet. In addition, the model generically predicts delayed variability at lower energies, associated with the recovery of the electron energy density once the blob exits the BLR and radiative cooling becomes less efficient. The combination of a fast $\gamma$-ray rise, a prolonged decay, and delayed or weak MWL counterparts thus provides a clear observational diagnostic for identifying flares primarily driven by bulk motion and geometrical effects, rather than by transient acceleration episodes. If the emission extends up to very-high energies during the flare, a delay would be expected with respect to the high-energy band.

Observationally, such asymmetric GeV flares with extended decay phases are not unique to the December 2013 event of 3C\,279, but have been reported in several other FSRQs. In particular, the orphan $\gamma$-ray flare of PKS\,1510-089 in April 2009, analysed by \citet{patel2021}, shows a clear fast-rise and slower-decay profile in the {\it Fermi}-LAT band, with no significant contemporaneous variability at lower energies. This event represents a close phenomenological analogue to the flare modelled here.

An asymmetric GeV flare with a comparable temporal structure, though accompanied by MWL activity (i.e.\ a non-orphan one), was observed in PKS\,1222+216 in June 2010 \citep{kushwaha2014mnras, kushwaha2014apj}, where the LAT light curve displays a relatively sharp rise followed by a more gradual decay. In that case, the decay time-scale was found to be difficult to reconcile with purely radiative losses, motivating interpretations involving dynamical effects within the jet. A similar (also non-orphan) $\gamma$-ray outburst with an extended multi-day decay has also been reported in PKS\,1502+106 during its 2008 high state \citep{abdo2010pks1502}. These examples indicate that asymmetric $\gamma$-ray flares with prolonged decay, including orphan events, are not isolated occurrences. The scenario explored in this work, in which the flare is governed by the bulk acceleration and displacement of a compact emitting region through stratified external radiation fields, provides a physically motivated framework that can, in principle, be applied to this broader class of events.

\section{Conclusion}
We have presented a two-zone leptonic scenario to interpret the
rapid orphan $\gamma$-ray flare exhibited by 3C\,279 on
20 December 2013. In contrast to models invoking sudden
variations of the particle injection or {\it ad hoc} acceleration episodes,
our approach relies solely on the bulk motion of a plasma blob
propagating through the external radiation fields of the AGN.

Using a time-dependent treatment of the particle evolution and
radiative processes in both a stationary and an accelerating blob,
we have shown that the flare naturally arises from the evolution of
the boosted external photon field as the moving blob crosses the
inner regions of the BLR. The combination of increasing Doppler
boost and enhanced external radiation energy density leads to a
short, intense EIC outburst with a hard spectrum and an
asymmetric light curve, consistent with observations. The lack of
optical variability is explained by the dominance of the stationary
blob in the synchrotron component.

Our model also predicts a delayed soft X-ray enhancement
associated with the gradual recovery of the electron energy
density once the blob exits the BLR, although no simultaneous
X-ray observations are available for this specific event.
If present, very-high-energy emission would also be delayed due
to $\gamma \gamma$ absorption. The
inferred energetics remain below the Eddington luminosity,
indicating that the mechanism is physically plausible.

These results suggest that bulk acceleration of emitting
structures within the acceleration-and-collimation zone of the jet
can account for a class of high-energy flares in FSRQs, including
orphan events. Future multiwavelength campaigns with
simultaneous X-ray and VHE coverage will be essential to test
the predicted delayed flares and to further constrain the dynamics
of plasma blobs within relativistic jets.

\section*{Acknowledgments}
The authors are grateful to C. Boisson for fruitful discussions and valuable insights throughout the development of this work, to K. Nalewajko for kindly providing published data points, and to the anonymous referee for insightful comments and suggestions.
AD acknowledges support from the Department of Science, Technology and Innovation, and the National Research Foundation of South Africa through the South African Gamma-Ray Astronomy Programme (SA-GAMMA).

\bibliographystyle{elsarticle-harv}

\begin{thebibliography}{60}
\expandafter\ifx\csname natexlab\endcsname\relax\def\natexlab#1{#1}\fi
\providecommand{\url}[1]{\texttt{#1}}
\providecommand{\href}[2]{#2}
\providecommand{\path}[1]{#1}
\providecommand{\DOIprefix}{doi:}
\providecommand{\ArXivprefix}{arXiv:}
\providecommand{\URLprefix}{URL: }
\providecommand{\Pubmedprefix}{pmid:}
\providecommand{\doi}[1]{\href{http://dx.doi.org/#1}{\path{#1}}}
\providecommand{\Pubmed}[1]{\href{pmid:#1}{\path{#1}}}
\providecommand{\bibinfo}[2]{#2}
\ifx\xfnm\relax \def\xfnm[#1]{\unskip,\space#1}\fi
\bibitem[{{Abdo} et~al.(2010){Abdo}, {Ackermann}, {Ajello}, {Atwood},
  {Axelsson}, {Baldini}, {Ballet}, {Barbiellini}, {Bastieri}, {Baughman},
  {Bechtol}, {Bellazzini}, {Berenji}, {Bloom}, {Bogaert}, {Bonamente},
  {Borgland}, {Bregeon}, {Brez}, {Brigida}, {Bruel}, {Burnett}, {Caliandro},
  {Cameron}, {Caraveo}, {Casandjian}, {Cavazzuti}, {Cecchi}, {{\c{C}}elik},
  {Chekhtman}, {Cheung}, {Chiang}, {Ciprini}, {Claus}, {Cohen-Tanugi},
  {Conrad}, {Cutini}, {Dermer}, {de Angelis}, {de Palma}, {Digel}, {Silva},
  {Drell}, {Dubois}, {Dumora}, {Farnier}, {Favuzzi}, {Fegan}, {Ferrara},
  {Focke}, {Frailis}, {Fuhrmann}, {Fukazawa}, {Funk}, {Fusco}, {Gargano},
  {Gasparrini}, {Gehrels}, {Germani}, {Giebels}, {Giglietto}, {Giordano},
  {Giroletti}, {Glanzman}, {Godfrey}, {Grenier}, {Grondin}, {Grove},
  {Guillemot}, {Guiriec}, {Hanabata}, {Harding}, {Hayashida}, {Hays}, {Hughes},
  {J{\'o}hannesson}, {Johnson}, {Johnson}, {Johnson}, {Kadler}, {Kamae},
  {Katagiri}, {Kataoka}, {Kerr}, {Kn{\"o}dlseder}, {Kocian}, {Kuehn}, {Kuss},
  {Lande}, {Latronico}, {Lemoine-Goumard}, {Longo}, {Loparco}, {Lott},
  {Lovellette}, {Lubrano}, {Madejski}, {Makeev}, {Marelli}, {Massaro},
  {Max-Moerbeck}, {Mazziotta}, {McConville}, {McEnery}, {Meurer}, {Michelson},
  {Mitthumsiri}, {Mizuno}, {Moiseev}, {Monte}, {Monzani}, {Morselli},
  {Moskalenko}, {Murgia}, {Nolan}, {Norris}, {Nuss}, {Ohsugi}, {Omodei},
  {Orlando}, {Ormes}, {Ozaki}, {Paneque}, {Panetta}, {Parent}, {Pavlidou},
  {Pearson}, {Pelassa}, {Pepe}, {Pesce-Rollins}, {Piron}, {Porter},
  {Rain{\`o}}, {Rando}, {Razzano}, {Razzaque}, {Readhead}, {Reimer}, {Reimer},
  {Reposeur}, {Richards}, {Ritz}, {Rochester}, {Rodriguez}, {Romani}, {Roth},
  {Ryde}, {Sadrozinski}, {Sanchez}, {Sander}, {Saz Parkinson}, {Scargle},
  {Sgr{\`o}}, {Shaw}, {Siskind}, {Smith}, {Smith}, {Spandre}, {Spinelli},
  {Stevenson}, {Strickman}, {Suson}, {Tajima}, {Takahashi}, {Tanaka}, {Thayer},
  {Thayer}, {Thompson}, {Tibaldo}, {Tibolla}, {Torres}, {Tosti}, {Tramacere},
  {Ubertini}, {Uchiyama}, {Usher}, {Vasileiou}, {Vilchez}, {Vitale}, {Waite},
  {Wang}, {Winer}, {Wood}, {Yasuda}, {Ylinen}, {Zensus}, {Ziegler}, {Fermi LAT
  Collaboration}, {Angelakis}, {Hovatta}, {Hoversten}, {Ikejiri}, {Kawabata},
  {Kovalev}, {Kovalev}, {Krichbaum}, {Lister}, {L{\"a}hteenm{\"a}ki},
  {Marchili} and {Ogle}}]{abdo2010pks1502}
\bibinfo{author}{{Abdo}, A.A.}, \bibinfo{author}{{Ackermann}, M.},
  \bibinfo{author}{{Ajello}, M.}, et~al.
\newblock , \bibinfo{year}{2010}.
\newblock \bibinfo{journal}{ApJ} \bibinfo{volume}{710},
  \bibinfo{pages}{810--827}.
\newblock \DOIprefix\doi{10.1088/0004-637X/710/1/810}.
\bibitem[{{Ackermann} et~al.(2016){Ackermann}, {Anantua}, {Asano}, {Baldini},
  {Barbiellini}, {Bastieri}, {Becerra Gonzalez}, {Bellazzini}, {Bissaldi},
  {Blandford}, {Bloom}, {Bonino}, {Bottacini}, {Bruel}, {Buehler}, {Caliandro},
  {Cameron}, {Caragiulo}, {Caraveo}, {Cavazzuti}, {Cecchi}, {Cheung}, {Chiang},
  {Chiaro}, {Ciprini}, {Cohen-Tanugi}, {Costanza}, {Cutini}, {D'Ammando}, {de
  Palma}, {Desiante}, {Digel}, {Di Lalla}, {Di Mauro}, {Di Venere}, {Drell},
  {Favuzzi}, {Fegan}, {Ferrara}, {Fukazawa}, {Funk}, {Fusco}, {Gargano},
  {Gasparrini}, {Giglietto}, {Giordano}, {Giroletti}, {Grenier}, {Guillemot},
  {Guiriec}, {Hayashida}, {Hays}, {Horan}, {J{\'o}hannesson}, {Kensei},
  {Kocevski}, {Kuss}, {La Mura}, {Larsson}, {Latronico}, {Li}, {Longo},
  {Loparco}, {Lott}, {Lovellette}, {Lubrano}, {Madejski}, {Magill}, {Maldera},
  {Manfreda}, {Mayer}, {Mazziotta}, {Michelson}, {Mirabal}, {Mizuno},
  {Monzani}, {Morselli}, {Moskalenko}, {Nalewajko}, {Negro}, {Nuss}, {Ohsugi},
  {Orlando}, {Paneque}, {Perkins}, {Pesce-Rollins}, {Piron}, {Pivato},
  {Porter}, {Principe}, {Rando}, {Razzano}, {Razzaque}, {Reimer}, {Scargle},
  {Sgr{\`o}}, {Sikora}, {Simone}, {Siskind}, {Spada}, {Spinelli}, {Stawarz},
  {Thayer}, {Thompson}, {Torres}, {Troja}, {Uchiyama}, {Yuan} and
  {Zimmer}}]{ackermann2016}
\bibinfo{author}{{Ackermann}, M.}, \bibinfo{author}{{Anantua}, R.},
  \bibinfo{author}{{Asano}, K.}, et~al.
\newblock , \bibinfo{year}{2016}.
\newblock \bibinfo{journal}{ApJl} \bibinfo{volume}{824}, \bibinfo{pages}{L20}.
\newblock \DOIprefix\doi{10.3847/2041-8205/824/2/L20}.
\bibitem[{{Asada} and {Nakamura}(2012)}]{2012ApJ...745L..28A}
\bibinfo{author}{{Asada}, K.}, \bibinfo{author}{{Nakamura}, M.}
\newblock , \bibinfo{year}{2012}.
\newblock \bibinfo{journal}{ApJl} \bibinfo{volume}{745}, \bibinfo{pages}{L28}.
\newblock \DOIprefix\doi{10.1088/2041-8205/745/2/L28}.
\bibitem[{{Asada} et~al.(2014){Asada}, {Nakamura}, {Doi}, {Nagai} and
  {Inoue}}]{2014ApJ...781L...2A}
\bibinfo{author}{{Asada}, K.}, \bibinfo{author}{{Nakamura}, M.},
  \bibinfo{author}{{Doi}, A.}, et~al.
\newblock , \bibinfo{year}{2014}.
\newblock \bibinfo{journal}{ApJl} \bibinfo{volume}{781}, \bibinfo{pages}{L2}.
\newblock \DOIprefix\doi{10.1088/2041-8205/781/1/L2}.
\bibitem[{{Baczko} et~al.(2022){Baczko}, {Ros}, {Kadler}, {Fromm}, {Boccardi},
  {Perucho}, {Krichbaum}, {Burd} and {Zensus}}]{2022A&A...658A.119B}
\bibinfo{author}{{Baczko}, A.K.}, \bibinfo{author}{{Ros}, E.},
  \bibinfo{author}{{Kadler}, M.}, et~al.
\newblock , \bibinfo{year}{2022}.
\newblock \bibinfo{journal}{A\&A} \bibinfo{volume}{658}, \bibinfo{pages}{A119}.
\newblock \DOIprefix\doi{10.1051/0004-6361/202141897}.
\bibitem[{Begelman et~al.(1984)Begelman, Blandford and
  Rees}]{RevModPhys.56.255}
\bibinfo{author}{Begelman, M.C.}, \bibinfo{author}{Blandford, R.D.},
  \bibinfo{author}{Rees, M.J.}
\newblock , \bibinfo{year}{1984}.
\newblock \bibinfo{journal}{Rev. Mod. Phys.} \bibinfo{volume}{56},
  \bibinfo{pages}{255--351}.
\newblock \DOIprefix\doi{10.1103/RevModPhys.56.255}.
\bibitem[{{Begelman} and {Li}(1994)}]{1994ApJ...426..269B}
\bibinfo{author}{{Begelman}, M.C.}, \bibinfo{author}{{Li}, Z.Y.}
\newblock , \bibinfo{year}{1994}.
\newblock \bibinfo{journal}{ApJ} \bibinfo{volume}{426}, \bibinfo{pages}{269}.
\newblock \DOIprefix\doi{10.1086/174061}.
\bibitem[{Beskin and Nokhrina(2006)}]{10.1111}
\bibinfo{author}{Beskin, V.S.}, \bibinfo{author}{Nokhrina, E.E.}
\newblock , \bibinfo{year}{2006}.
\newblock \bibinfo{journal}{MNRAS} \bibinfo{volume}{367},
  \bibinfo{pages}{375--386}.
\newblock \DOIprefix\doi{10.1111/j.1365-2966.2006.09957.x}.
\bibitem[{{Bloom} et~al.(2013){Bloom}, {Fromm} and {Ros}}]{Bloom2013}
\bibinfo{author}{{Bloom}, S.D.}, \bibinfo{author}{{Fromm}, C.M.},
  \bibinfo{author}{{Ros}, E.}
\newblock , \bibinfo{year}{2013}.
\newblock \bibinfo{journal}{AJ} \bibinfo{volume}{145}, \bibinfo{pages}{12}.
\newblock \DOIprefix\doi{10.1088/0004-6256/145/1/12}.
\bibitem[{{Burbidge} and {Rosenberg}(1965)}]{redshift}
\bibinfo{author}{{Burbidge}, E.M.}, \bibinfo{author}{{Rosenberg}, F.D.}
\newblock , \bibinfo{year}{1965}.
\newblock \bibinfo{journal}{ApJ} \bibinfo{volume}{142}, \bibinfo{pages}{1673}.
\newblock \DOIprefix\doi{10.1086/148458}.
\bibitem[{{Burd} et~al.(2022){Burd}, {Kadler}, {Mannheim}, {Baczko}, {Ringholz}
  and {Ros}}]{Burd2022}
\bibinfo{author}{{Burd}, P.R.}, \bibinfo{author}{{Kadler}, M.},
  \bibinfo{author}{{Mannheim}, K.}, et~al.
\newblock , \bibinfo{year}{2022}.
\newblock \bibinfo{journal}{A\&A} \bibinfo{volume}{660}, \bibinfo{pages}{A1}.
\newblock \DOIprefix\doi{10.1051/0004-6361/202142363}.
\bibitem[{Bottcher and Els(2016)}]{Bottcher_2016}
\bibinfo{author}{Bottcher, M.}, \bibinfo{author}{Els, P.}
\newblock , \bibinfo{year}{2016}.
\newblock \bibinfo{journal}{AJ} \bibinfo{volume}{821}, \bibinfo{pages}{102}.
\newblock \DOIprefix\doi{10.3847/0004-637X/821/2/102}.
\bibitem[{{Calderone} et~al.(2012){Calderone}, {Sbarrato} and {Ghisellini}}]{frac_DT}
\bibinfo{author}{{Calderone}, G.}, \bibinfo{author}{{Sbarrato}, T.},
  \bibinfo{author}{{Ghisellini}, G.}
\newblock , \bibinfo{year}{2012}.
\newblock \bibinfo{journal}{MNRAS} \bibinfo{volume}{425},
  \bibinfo{pages}{L41--L45}.
\newblock \DOIprefix\doi{10.1111/j.1745-3933.2012.01296.x}.
\bibitem[{{Camenzind}(1987)}]{1987A&A...184..341C}
\bibinfo{author}{{Camenzind}, M.}
\newblock , \bibinfo{year}{1987}.
\newblock \bibinfo{journal}{A\&A} \bibinfo{volume}{184},
  \bibinfo{pages}{341--360}.
\bibitem[{Dermer et~al.(2014)Dermer, Cerruti, Lott, Boisson and
  Zech}]{dermer_equipartition_2014}
\bibinfo{author}{Dermer, C.D.}, \bibinfo{author}{Cerruti, M.},
  \bibinfo{author}{Lott, B.}, et~al.
\newblock , \bibinfo{year}{2014}.
\newblock \bibinfo{journal}{ApJ} \bibinfo{volume}{782}, \bibinfo{pages}{82}.
\bibitem[{{Dmytriiev} and {B{\"o}ttcher}(2024)}]{dmytriiev2024}
\bibinfo{author}{{Dmytriiev}, A.}, \bibinfo{author}{{B{\"o}ttcher}, M.}
\newblock , \bibinfo{year}{2024}.
\newblock \bibinfo{journal}{A\&A} \bibinfo{volume}{687}, \bibinfo{pages}{A64}.
\newblock \DOIprefix\doi{10.1051/0004-6361/202348269}.
\bibitem[{{Dmytriiev} et~al.(2023){Dmytriiev}, {B{\"o}ttcher} and
  {Machipi}}]{dmytriiev2023}
\bibinfo{author}{{Dmytriiev}, A.}, \bibinfo{author}{{B{\"o}ttcher}, M.},
  \bibinfo{author}{{Machipi}, T.O.}
\newblock , \bibinfo{year}{2023}.
\newblock \bibinfo{journal}{ApJ} \bibinfo{volume}{949}, \bibinfo{pages}{28}.
\newblock \DOIprefix\doi{10.3847/1538-4357/acc57b}.
\bibitem[{{Dmytriiev} et~al.(2021){Dmytriiev}, {Sol} and
  {Zech}}]{dmytriiev2021}
\bibinfo{author}{{Dmytriiev}, A.}, \bibinfo{author}{{Sol}, H.},
  \bibinfo{author}{{Zech}, A.}
\newblock , \bibinfo{year}{2021}.
\newblock \bibinfo{journal}{MNRAS} \bibinfo{volume}{505},
  \bibinfo{pages}{2712--2730}.
\newblock \DOIprefix\doi{10.1093/MNRAS/stab1445}.
\bibitem[{{Dom{\'\i}nguez} et~al.(2011){Dom{\'\i}nguez}, {Primack}, {Rosario},
  {Prada}, {Gilmore}, {Faber}, {Koo}, {Somerville}, {P{\'e}rez-Torres},
  {P{\'e}rez-Gonz{\'a}lez}, {Huang}, {Davis}, {Guhathakurta}, {Barmby},
  {Conselice}, {Lozano}, {Newman} and {Cooper}}]{dominguez2011}
\bibinfo{author}{{Dom{\'\i}nguez}, A.}, \bibinfo{author}{{Primack}, J.R.},
  \bibinfo{author}{{Rosario}, D.J.}, et~al.
\newblock , \bibinfo{year}{2011}.
\newblock \bibinfo{journal}{MNRAS} \bibinfo{volume}{410},
  \bibinfo{pages}{2556--2578}.
\newblock \DOIprefix\doi{10.1111/j.1365-2966.2010.17631.x}.
\bibitem[{{Finke}(2016)}]{Finke_2016}
\bibinfo{author}{{Finke}, J.D.}
\newblock , \bibinfo{year}{2016}.
\newblock \bibinfo{journal}{ApJ} \bibinfo{volume}{830}, \bibinfo{pages}{94}.
\newblock \DOIprefix\doi{10.3847/0004-637X/830/2/94}.
\bibitem[{{Ghisellini} and {Tavecchio}(2009)}]{ghisellini2009}
\bibinfo{author}{{Ghisellini}, G.}, \bibinfo{author}{{Tavecchio}, F.}
\newblock , \bibinfo{year}{2009}.
\newblock \bibinfo{journal}{MNRAS} \bibinfo{volume}{397},
  \bibinfo{pages}{985--1002}.
\newblock \DOIprefix\doi{10.1111/j.1365-2966.2009.15007.x}.
\bibitem[{{H.~E.~S.~S. Collaboration} et~al.(2019){H.~E.~S.~S. Collaboration},
  {Abdalla}, {Adam}, {Aharonian}, {Ait Benkhali}, {Ang{\"u}ner}, {Arakawa},
  {Arcaro}, {Armand}, {Ashkar}, {Backes}, {Barbosa Martins}, {Barnard},
  {Becherini}, {Berge}, {Bernl{\"o}hr}, {Blackwell}, {B{\"o}ttcher}, {Boisson},
  {Bolmont}, {Bonnefoy}, {Bregeon}, {Breuhaus}, {Brun}, {Brun}, {Bryan},
  {B{\"u}chele}, {Bulik}, {Bylund}, {Capasso}, {Caroff}, {Carosi}, {Casanova},
  {Cerruti}, {Chand}, {Chandra}, {Chen}, {Colafrancesco}, {Cury{\l}o},
  {Davids}, {Deil}, {Devin}, {deWilt}, {Dirson}, {Djannati-Ata{\"\i}},
  {Dmytriiev}, {Donath}, {Doroshenko}, {Drury}, {Dyks}, {Egberts}, {Emery},
  {Ernenwein}, {Eschbach}, {Feijen}, {Fegan}, {Fiasson}, {Fontaine}, {Funk},
  {F{\"u}{\ss}ling}, {Gabici}, {Gallant}, {Gat{\'e}}, {Giavitto}, {Glawion},
  {Glicenstein}, {Gottschall}, {Grondin}, {Hahn}, {Haupt}, {Heinzelmann},
  {Henri}, {Hermann}, {Hinton}, {Hofmann}, {Hoischen}, {Holch}, {Holler},
  {Horns}, {Huber}, {Iwasaki}, {Jamrozy}, {Jankowsky}, {Jankowsky},
  {Jardin-Blicq}, {Jung-Richardt}, {Kastendieck}, {Katarzy{\'n}ski},
  {Katsuragawa}, {Katz}, {Khangulyan}, {Kh{\'e}lifi}, {King}, {Klepser},
  {Klu{\'z}niak}, {Komin}, {Kosack}, {Kostunin}, {Kraus}, {Lamanna}, {Lau},
  {Lemi{\`e}re}, {Lemoine-Goumard}, {Lenain}, {Leser}, {Levy}, {Lohse},
  {Lypova}, {Mackey}, {Majumdar}, {Malyshev}, {Marandon}, {Marcowith}, {Mares},
  {Mariaud}, {Mart{\'\i}-Devesa}, {Marx}, {Maurin}, {Meintjes}, {Mitchell},
  {Moderski}, {Mohamed}, {Mohrmann}, {Moore}, {Moulin}, {Muller}, {Murach},
  {Nakashima}, {de Naurois}, {Ndiyavala}, {Niederwanger}, {Niemiec}, {Oakes},
  {O'Brien}, {Odaka}, {Ohm}, {de Ona Wilhelmi}, {Ostrowski}, {Oya}, {Panter},
  {Parsons}, {Perennes}, {Petrucci}, {Peyaud}, {Piel}, {Pita}, {Poireau},
  {Priyana Noel}, {Prokhorov}, {Prokoph}, {P{\"u}hlhofer}, {Punch},
  {Quirrenbach}, {Raab}, {Rauth}, {Reimer}, {Reimer}, {Remy}, {Renaud},
  {Rieger}, {Rinchiuso}, {Romoli}, {Rowell}, {Rudak}, {Ruiz-Velasco},
  {Sahakian}, {Saito}, {Sanchez}, {Santangelo}, {Sasaki}, {Schlickeiser},
  {Sch{\"u}ssler}, {Schulz}, {Schutte}, {Schwanke}, {Schwemmer},
  {Seglar-Arroyo}, {Senniappan}, {Seyffert}, {Shafi}, {Shiningayamwe},
  {Simoni}, {Sinha}, {Sol}, {Specovius}, {Spir-Jacob}, {Stawarz}, {Steenkamp},
  {Stegmann}, {Steppa}, {Takahashi}, {Tavernier}, {Taylor}, {Terrier},
  {Tiziani}, {Tluczykont}, {Trichard}, {Tsirou}, {Tsuji} and
  {Tuffs}}]{hess2019}
\bibinfo{author}{{H.~E.~S.~S. Collaboration}}, \bibinfo{author}{{Abdalla}, H.},
  \bibinfo{author}{{Adam}, R.}, et~al.
\newblock , \bibinfo{year}{2019}.
\newblock \bibinfo{journal}{A\&A} \bibinfo{volume}{627}, \bibinfo{pages}{A159}.
\newblock \DOIprefix\doi{10.1051/0004-6361/201935704}.
\bibitem[{Hada et~al.(2018)Hada, Doi, Wajima, D‚ÄôAmmando, Orienti, Giroletti,
  Giovannini, Nakamura and Asada}]{Hada_2018}
\bibinfo{author}{Hada, K.}, \bibinfo{author}{Doi, A.}, \bibinfo{author}{Wajima,
  K.}, et~al.
\newblock , \bibinfo{year}{2018}.
\newblock \bibinfo{journal}{AJ} \bibinfo{volume}{860}, \bibinfo{pages}{141}.
\newblock \DOIprefix\doi{10.3847/1538-4357/aac49f}.
\bibitem[{{Hayashida} et~al.(2012){Hayashida}, {Madejski}, {Nalewajko},
  {Sikora}, {Wehrle}, {Ogle}, {Collmar}, {Larsson}, {Fukazawa}, {Itoh},
  {Chiang}, {Stawarz}, {Blandford}, {Richards}, {Max-Moerbeck}, {Readhead},
  {Buehler}, {Cavazzuti}, {Ciprini}, {Gehrels}, {Reimer}, {Szostek}, {Tanaka},
  {Tosti}, {Uchiyama}, {Kawabata}, {Kino}, {Sakimoto}, {Sasada}, {Sato},
  {Uemura}, {Yamanaka}, {Greiner}, {Kruehler}, {Rossi}, {Macquart}, {Bock},
  {Villata}, {Raiteri}, {Agudo}, {Aller}, {Aller}, {Arkharov}, {Bach},
  {Ben{\'\i}tez}, {Berdyugin}, {Blinov}, {Blumenthal}, {B{\"o}ttcher}, {Buemi},
  {Carosati}, {Chen}, {Di Paola}, {Dolci}, {Efimova}, {Forn{\'e}}, {G{\'o}mez},
  {Gurwell}, {Heidt}, {Hiriart}, {Jordan}, {Jorstad}, {Joshi}, {Kimeridze},
  {Konstantinova}, {Kopatskaya}, {Koptelova}, {Kurtanidze},
  {L{\"a}hteenm{\"a}ki}, {Lamerato}, {Larionov}, {Larionova}, {Larionova},
  {Leto}, {Lindfors}, {Marscher}, {McHardy}, {Molina}, {Morozova},
  {Nikolashvili}, {Nilsson}, {Reinthal}, {Roustazadeh}, {Sakamoto}, {Sigua},
  {Sillanp{\"a}{\"a}}, {Takalo}, {Tammi}, {Taylor}, {Tornikoski}, {Trigilio},
  {Troitsky} and {Umana}}]{hayashida2012}
\bibinfo{author}{{Hayashida}, M.}, \bibinfo{author}{{Madejski}, G.M.},
  \bibinfo{author}{{Nalewajko}, K.}, et~al.
\newblock , \bibinfo{year}{2012}.
\newblock \bibinfo{journal}{ApJ} \bibinfo{volume}{754}, \bibinfo{pages}{114}.
\newblock \DOIprefix\doi{10.1088/0004-637X/754/2/114}.
\bibitem[{{Hayashida} et~al.(2015){Hayashida}, {Nalewajko}, {Madejski},
  {Sikora}, {Itoh}, {Ajello}, {Blandford}, {Buson}, {Chiang}, {Fukazawa},
  {Furniss}, {Urry}, {Hasan}, {Harrison}, {Alexander}, {Balokovi{\'c}},
  {Barret}, {Boggs}, {Christensen}, {Craig}, {Forster}, {Giommi},
  {Grefenstette}, {Hailey}, {Hornstrup}, {Kitaguchi}, {Koglin}, {Madsen},
  {Mao}, {Miyasaka}, {Mori}, {Perri}, {Pivovaroff}, {Puccetti}, {Rana},
  {Stern}, {Tagliaferri}, {Westergaard}, {Zhang}, {Zoglauer}, {Gurwell},
  {Uemura}, {Akitaya}, {Kawabata}, {Kawaguchi}, {Kanda}, {Moritani}, {Takaki},
  {Ui}, {Yoshida}, {Agarwal} and {Gupta}}]{Hayashida_2015}
\bibinfo{author}{{Hayashida}, M.}, \bibinfo{author}{{Nalewajko}, K.},
  \bibinfo{author}{{Madejski}, G.M.}, et~al.
\newblock , \bibinfo{year}{2015}.
\newblock \bibinfo{journal}{ApJ} \bibinfo{volume}{807}, \bibinfo{pages}{79}.
\newblock \DOIprefix\doi{10.1088/0004-637X/807/1/79}.
\bibitem[{{Homan} et~al.(2009){Homan}, {Kadler}, {Kellermann}, {Kovalev},
  {Lister}, {Ros}, {Savolainen} and {Zensus}}]{Homan2009}
\bibinfo{author}{{Homan}, D.C.}, \bibinfo{author}{{Kadler}, M.},
  \bibinfo{author}{{Kellermann}, K.I.}, et~al.
\newblock , \bibinfo{year}{2009}.
\newblock \bibinfo{journal}{ApJ} \bibinfo{volume}{706},
  \bibinfo{pages}{1253--1268}.
\newblock \DOIprefix\doi{10.1088/0004-637X/706/2/1253}.
\bibitem[{{Jorstad} et~al.(2017){Jorstad}, {Marscher}, {Morozova}, {Troitsky},
  {Agudo}, {Casadio}, {Foord}, {G{\'o}mez}, {MacDonald}, {Molina},
  {L{\"a}hteenm{\"a}ki}, {Tammi} and {Tornikoski}}]{Jorstad2017}
\bibinfo{author}{{Jorstad}, S.G.}, \bibinfo{author}{{Marscher}, A.P.},
  \bibinfo{author}{{Morozova}, D.A.}, et~al.
\newblock , \bibinfo{year}{2017}.
\newblock \bibinfo{journal}{ApJ} \bibinfo{volume}{846}, \bibinfo{pages}{98}.
\newblock \DOIprefix\doi{10.3847/1538-4357/aa8407}.
\bibitem[{{Kardashev}(1962)}]{kardashev1962}
\bibinfo{author}{{Kardashev}, N.S.}
\newblock , \bibinfo{year}{1962}.
\newblock \bibinfo{journal}{Sov. Astron.} \bibinfo{volume}{6},
  \bibinfo{pages}{317}.
\bibitem[{{Kiehlmann} et~al.(2016){Kiehlmann}, {Savolainen}, {Jorstad},
  {Sokolovsky}, {Schinzel}, {Marscher}, {Larionov}, {Agudo}, {Akitaya},
  {Ben{\'\i}tez}, {Berdyugin}, {Blinov}, {Bochkarev}, {Borman}, {Burenkov},
  {Casadio}, {Doroshenko}, {Efimova}, {Fukazawa}, {G{\'o}mez}, {Grishina},
  {Hagen-Thorn}, {Heidt}, {Hiriart}, {Itoh}, {Joshi}, {Kawabata}, {Kimeridze},
  {Kopatskaya}, {Korobtsev}, {Krajci}, {Kurtanidze}, {Kurtanidze}, {Larionova},
  {Larionova}, {Lindfors}, {L{\'o}pez}, {McHardy}, {Molina}, {Moritani},
  {Morozova}, {Nazarov}, {Nikolashvili}, {Nilsson}, {Pulatova}, {Reinthal},
  {Sadun}, {Sasada}, {Savchenko}, {Sergeev}, {Sigua}, {Smith}, {Sorcia},
  {Spiridonova}, {Takaki}, {Takalo}, {Taylor}, {Troitsky}, {Uemura},
  {Ugolkova}, {Ui}, {Yoshida}, {Zensus} and {Zhdanova}}]{kiehlmann2016}
\bibinfo{author}{{Kiehlmann}, S.}, \bibinfo{author}{{Savolainen}, T.},
  \bibinfo{author}{{Jorstad}, S.G.}, et~al.
\newblock , \bibinfo{year}{2016}.
\newblock \bibinfo{journal}{A\&A} \bibinfo{volume}{590}, \bibinfo{pages}{A10}.
\newblock \DOIprefix\doi{10.1051/0004-6361/201527725}.
\bibitem[{{Kim} et~al.(2020){Kim}, {Krichbaum}, {Broderick}, {Wielgus},
  {Blackburn}, {G{\'o}mez}, {Johnson}, {Bouman}, {Chael}, {Akiyama}, {Jorstad},
  {Marscher}, {Issaoun}, {Janssen}, {Chan}, {Savolainen}, {Pesce}, {{\"O}zel},
  {Alberdi}, {Alef}, {Asada}, {Azulay}, {Baczko}, {Ball}, {Balokovi{\'c}},
  {Barrett}, {Bintley}, {Boland}, {Bower}, {Bremer}, {Brinkerink},
  {Brissenden}, {Britzen}, {Broguiere}, {Bronzwaer}, {Byun}, {Carlstrom},
  {Chatterjee}, {Chatterjee}, {Chen}, {Chen}, {Cho}, {Christian}, {Conway},
  {Cordes}, {Crew}, {Cui}, {Davelaar}, {De Laurentis}, {Deane}, {Dempsey},
  {Desvignes}, {Dexter}, {Doeleman}, {Eatough}, {Falcke}, {Fish}, {Fomalont},
  {Fraga-Encinas}, {Friberg}, {Fromm}, {Galison}, {Gammie}, {Garc{\'\i}a},
  {Gentaz}, {Georgiev}, {Goddi}, {Gold}, {G{\'o}mez-Ruiz}, {Gu}, {Gurwell},
  {Hada}, {Hecht}, {Hesper}, {Ho}, {Ho}, {Honma}, {Huang}, {Huang}, {Hughes},
  {Ikeda}, {Inoue}, {James}, {Jannuzi}, {Jeter}, {Jiang}, {Jimenez-Rosales},
  {Jung}, {Karami}, {Karuppusamy}, {Kawashima}, {Keating}, {Kettenis}, {Kim},
  {Kim}, {Kino}, {Koay}, {Koch}, {Koyama}, {Kramer}, {Kramer}, {Kuo}, {Lauer},
  {Lee}, {Li}, {Li}, {Lindqvist}, {Lico}, {Liu}, {Liuzzo}, {Lo}, {Lobanov},
  {Loinard}, {Lonsdale}, {Lu}, {MacDonald}, {Mao}, {Markoff}, {Marrone},
  {Mart{\'\i}-Vidal}, {Matsushita}, {Matthews}, {Medeiros}, {Menten}, {Mizuno},
  {Mizuno}, {Moran}, {Moriyama}, {Moscibrodzka}, {Musoke}, {M{\"u}ller},
  {Nagai}, {Nagar}, {Nakamura}, {Narayan}, {Narayanan}, {Natarajan}, {Neri},
  {Ni}, {Noutsos}, {Okino}, {Olivares}, {Ortiz-Le{\'o}n}, {Oyama}, {Palumbo},
  {Park}, {Patel}, {Pen}, {Pi{\'e}tu}, {Plambeck}, {PopStefanija}, {Porth},
  {Prather}, {Preciado-L{\'o}pez}, {Psaltis}, {Pu}, {Ramakrishnan}, {Rao},
  {Rawlings}, {Raymond}, {Rezzolla}, {Ripperda}, {Roelofs}, {Rogers}, {Ros},
  {Rose}, {Roshanineshat}, {Rottmann}, {Roy}, {Ruszczyk}, {Ryan}, {Rygl},
  {S{\'a}nchez}, {S{\'a}nchez-Arguelles}, {Sasada}, {Schloerb}, {Schuster},
  {Shao}, {Shen}, {Small}, {Sohn}, {SooHoo}, {Tazaki}, {Tiede}, {Tilanus},
  {Titus}, {Toma}, {Torne}, {Trent}, {Traianou}, {Trippe}, {Tsuda}, {van
  Bemmel}, {van Langevelde}, {van Rossum}, {Wagner}, {Wardle}, {Ward-Thompson},
  {Weintroub} and {Wex}}]{Kim2020}
\bibinfo{author}{{Kim}, J.Y.}, \bibinfo{author}{{Krichbaum}, T.P.},
  \bibinfo{author}{{Broderick}, A.E.}, et~al.
\newblock , \bibinfo{year}{2020}.
\newblock \bibinfo{journal}{A\&A} \bibinfo{volume}{640}, \bibinfo{pages}{A69}.
\newblock \DOIprefix\doi{10.1051/0004-6361/202037493}.
\bibitem[{{Komissarov}(2011)}]{komissarov_theory}
\bibinfo{author}{{Komissarov}, S.S.}
\newblock , \bibinfo{year}{2011}.
\newblock \bibinfo{journal}{Mem. Soc. Astron. Ital.} \bibinfo{volume}{82},
  \bibinfo{pages}{95}.
\newblock \DOIprefix\doi{10.48550/arXiv.1006.2242}.
\bibitem[{{Komissarov} et~al.(2009){Komissarov}, {Vlahakis}, {K{\"o}nigl} and
  {Barkov}}]{komissarov_num}
\bibinfo{author}{{Komissarov}, S.S.}, \bibinfo{author}{{Vlahakis}, N.},
  \bibinfo{author}{{K{\"o}nigl}, A.}, et~al.
\newblock , \bibinfo{year}{2009}.
\newblock \bibinfo{journal}{MNRAS} \bibinfo{volume}{394},
  \bibinfo{pages}{1182--1212}.
\newblock \DOIprefix\doi{10.1111/j.1365-2966.2009.14410.x}.
\bibitem[{{Kushwaha} et~al.(2014a){Kushwaha}, {Sahayanathan}, {Lekshmi},
  {Singh}, {Bhattacharyya} and {Bhattacharya}}]{kushwaha2014mnras}
\bibinfo{author}{{Kushwaha}, P.}, \bibinfo{author}{{Sahayanathan}, S.},
  \bibinfo{author}{{Lekshmi}, R.}, et~al.
\newblock , \bibinfo{year}{2014}a.
\newblock \bibinfo{journal}{MNRAS} \bibinfo{volume}{442},
  \bibinfo{pages}{131--137}.
\newblock \DOIprefix\doi{10.1093/mnras/stu836}.
\bibitem[{{Kushwaha} et~al.(2014b){Kushwaha}, {Singh} and
  {Sahayanathan}}]{kushwaha2014apj}
\bibinfo{author}{{Kushwaha}, P.}, \bibinfo{author}{{Singh}, K.P.},
  \bibinfo{author}{{Sahayanathan}, S.}
\newblock , \bibinfo{year}{2014}b.
\newblock \bibinfo{journal}{ApJ} \bibinfo{volume}{796}, \bibinfo{pages}{61}.
\newblock \DOIprefix\doi{10.1088/0004-637X/796/1/61}.
\bibitem[{{Lewis} et~al.(2019){Lewis}, {Finke} and {Becker}}]{Lewis2019}
\bibinfo{author}{{Lewis}, T.R.}, \bibinfo{author}{{Finke}, J.D.},
  \bibinfo{author}{{Becker}, P.A.}
\newblock , \bibinfo{year}{2019}.
\newblock \bibinfo{journal}{ApJ} \bibinfo{volume}{884}, \bibinfo{pages}{116}.
\newblock \DOIprefix\doi{10.3847/1538-4357/ab43c3}.
\bibitem[{{Li} et~al.(1992){Li}, {Chiueh} and {Begelman}}]{1992ApJ...394..459L}
\bibinfo{author}{{Li}, Z.Y.}, \bibinfo{author}{{Chiueh}, T.},
  \bibinfo{author}{{Begelman}, M.C.}
\newblock , \bibinfo{year}{1992}.
\newblock \bibinfo{journal}{ApJ} \bibinfo{volume}{394}, \bibinfo{pages}{459}.
\newblock \DOIprefix\doi{10.1086/171597}.
\bibitem[{{Lister} et~al.(2018){Lister}, {Aller}, {Aller}, {Hodge}, {Homan},
  {Kovalev}, {Pushkarev} and {Savolainen}}]{Lister2018}
\bibinfo{author}{{Lister}, M.L.}, \bibinfo{author}{{Aller}, M.F.},
  \bibinfo{author}{{Aller}, H.D.}, et~al.
\newblock , \bibinfo{year}{2018}.
\newblock \bibinfo{journal}{ApJs} \bibinfo{volume}{234}, \bibinfo{pages}{12}.
\newblock \DOIprefix\doi{10.3847/1538-4365/aa9c44}.
\bibitem[{{Lister} et~al.(2016){Lister}, {Aller}, {Aller}, {Homan},
  {Kellermann}, {Kovalev}, {Pushkarev}, {Richards}, {Ros} and
  {Savolainen}}]{Lister2016}
\bibinfo{author}{{Lister}, M.L.}, \bibinfo{author}{{Aller}, M.F.},
  \bibinfo{author}{{Aller}, H.D.}, et~al.
\newblock , \bibinfo{year}{2016}.
\newblock \bibinfo{journal}{AJ} \bibinfo{volume}{152}, \bibinfo{pages}{12}.
\newblock \DOIprefix\doi{10.3847/0004-6256/152/1/12}.
\bibitem[{{Lister} et~al.(2021){Lister}, {Homan}, {Kellermann}, {Kovalev},
  {Pushkarev}, {Ros} and {Savolainen}}]{Lister2021}
\bibinfo{author}{{Lister}, M.L.}, \bibinfo{author}{{Homan}, D.C.},
  \bibinfo{author}{{Kellermann}, K.I.}, et~al.
\newblock , \bibinfo{year}{2021}.
\newblock \bibinfo{journal}{ApJ} \bibinfo{volume}{923}, \bibinfo{pages}{30}.
\newblock \DOIprefix\doi{10.3847/1538-4357/ac230f}.
\bibitem[{Lyubarsky(2009)}]{Lyubarsky_2009}
\bibinfo{author}{Lyubarsky, Y.}
\newblock , \bibinfo{year}{2009}.
\newblock \bibinfo{journal}{AJ} \bibinfo{volume}{698}, \bibinfo{pages}{1570}.
\newblock \DOIprefix\doi{10.1088/0004-637X/698/2/1570}.
\bibitem[{{MAGIC Collaboration} et~al.(2008){MAGIC Collaboration}, {Albert},
  {Aliu}, {Anderhub}, {Antonelli}, {Antoranz}, {Backes}, {Baixeras}, {Barrio},
  {Bartko}, {Bastieri}, {Becker}, {Bednarek}, {Berger}, {Bernardini},
  {Bigongiari}, {Biland}, {Bock}, {Bonnoli}, {Bordas}, {Bosch-Ramon}, {Bretz},
  {Britvitch}, {Camara}, {Carmona}, {Chilingarian}, {Commichau}, {Contreras},
  {Cortina}, {Costado}, {Covino}, {Curtef}, {Dazzi}, {De Angelis}, {de Cea del
  Pozo}, {de los Reyes}, {De Lotto}, {De Maria}, {De Sabata}, {Delgado Mendez},
  {Dominguez}, {Dorner}, {Doro}, {Errando}, {Fagiolini}, {Ferenc},
  {Fern{\'a}ndez}, {Firpo}, {Fonseca}, {Font}, {Galante}, {Garc{\'\i}a
  L{\'o}pez}, {Garczarczyk}, {Gaug}, {Goebel}, {Hayashida}, {Herrero},
  {H{\"o}hne}, {Hose}, {Hsu}, {Huber}, {Jogler}, {Kneiske}, {Kranich}, {La
  Barbera}, {Laille}, {Leonardo}, {Lindfors}, {Lombardi}, {Longo}, {L{\'o}pez},
  {Lorenz}, {Majumdar}, {Maneva}, {Mankuzhiyil}, {Mannheim}, {Maraschi},
  {Mariotti}, {Mart{\'\i}nez}, {Mazin}, {Meucci}, {Meyer}, {Miranda},
  {Mirzoyan}, {Mizobuchi}, {Moles}, {Moralejo}, {Nieto}, {Nilsson}, {Ninkovic},
  {Otte}, {Oya}, {Panniello}, {Paoletti}, {Paredes}, {Pasanen}, {Pascoli},
  {Pauss}, {Pegna}, {Perez-Torres}, {Persic}, {Peruzzo}, {Piccioli}, {Prada},
  {Prandini}, {Puchades}, {Raymers}, {Rhode}, {Rib{\'o}}, {Rico}, {Rissi},
  {Robert}, {R{\"u}gamer}, {Saggion}, {Saito}, {Salvati}, {Sanchez-Conde},
  {Sartori}, {Satalecka}, {Scalzotto}, {Scapin}, {Schmitt}, {Schweizer},
  {Shayduk}, {Shinozaki}, {Shore}, {Sidro}, {Sierpowska-Bartosik},
  {Sillanp{\"a}{\"a}}, {Sobczynska}, {Spanier}, {Stamerra}, {Stark}, {Takalo},
  {Tavecchio}, {Temnikov}, {Tescaro}, {Teshima}, {Tluczykont}, {Torres},
  {Turini}, {Vankov}, {Venturini}, {Vitale}, {Wagner}, {Wittek}, {Zabalza},
  {Zandanel}, {Zanin} and {Zapatero}}]{magic2008}
\bibinfo{author}{{MAGIC Collaboration}}, \bibinfo{author}{{Albert}, J.},
  \bibinfo{author}{{Aliu}, E.}, et~al.
\newblock , \bibinfo{year}{2008}.
\newblock \bibinfo{journal}{Science} \bibinfo{volume}{320},
  \bibinfo{pages}{1752}.
\newblock \DOIprefix\doi{10.1126/science.1157087}.
\bibitem[{{Nakamura} and {Asada}(2013)}]{2013ApJ...775..118N}
\bibinfo{author}{{Nakamura}, M.}, \bibinfo{author}{{Asada}, K.}
\newblock , \bibinfo{year}{2013}.
\newblock \bibinfo{journal}{ApJ} \bibinfo{volume}{775}, \bibinfo{pages}{118}.
\newblock \DOIprefix\doi{10.1088/0004-637X/775/2/118}.
\bibitem[{{Nilsson} et~al.(2009){Nilsson}, {Pursimo}, {Villforth}, {Lindfors}
  and {Takalo}}]{massbh}
\bibinfo{author}{{Nilsson}, K.}, \bibinfo{author}{{Pursimo}, T.},
  \bibinfo{author}{{Villforth}, C.}, et~al.
\newblock , \bibinfo{year}{2009}.
\newblock \bibinfo{journal}{A\&A} \bibinfo{volume}{505},
  \bibinfo{pages}{601--604}.
\newblock \DOIprefix\doi{10.1051/0004-6361/200912820}.
\bibitem[{{Paliya} et~al.(2016){Paliya}, {Diltz}, {B{\"o}ttcher}, {Stalin} and
  {Buckley}}]{Paliya2016}
\bibinfo{author}{{Paliya}, V.S.}, \bibinfo{author}{{Diltz}, C.},
  \bibinfo{author}{{B{\"o}ttcher}, M.}, et~al.
\newblock , \bibinfo{year}{2016}.
\newblock \bibinfo{journal}{ApJ} \bibinfo{volume}{817}, \bibinfo{pages}{61}.
\newblock \DOIprefix\doi{10.3847/0004-637X/817/1/61}.
\bibitem[{{Park} et~al.(2021){Park}, {Hada}, {Nakamura}, {Asada}, {Zhao} and
  {Kino}}]{2021ApJ...909...76P}
\bibinfo{author}{{Park}, J.}, \bibinfo{author}{{Hada}, K.},
  \bibinfo{author}{{Nakamura}, M.}, et~al.
\newblock , \bibinfo{year}{2021}.
\newblock \bibinfo{journal}{ApJ} \bibinfo{volume}{909}, \bibinfo{pages}{76}.
\newblock \DOIprefix\doi{10.3847/1538-4357/abd6ee}.
\bibitem[{{Patel} et~al.(2021){Patel}, {Bose}, {Gupta} and
  {Zuberi}}]{patel2021}
\bibinfo{author}{{Patel}, S.R.}, \bibinfo{author}{{Bose}, D.},
  \bibinfo{author}{{Gupta}, N.}, et~al.
\newblock , \bibinfo{year}{2021}.
\newblock \bibinfo{journal}{Journal of High Energy Astrophysics}
  \bibinfo{volume}{29}, \bibinfo{pages}{31--39}.
\newblock \DOIprefix\doi{10.1016/j.jheap.2020.12.001}.
\bibitem[{{Pushkarev} et~al.(2012){Pushkarev}, {Hovatta}, {Kovalev}, {Lister},
  {Lobanov}, {Savolainen} and {Zensus}}]{Pushkarev2012}
\bibinfo{author}{{Pushkarev}, A.B.}, \bibinfo{author}{{Hovatta}, T.},
  \bibinfo{author}{{Kovalev}, Y.Y.}, et~al.
\newblock , \bibinfo{year}{2012}.
\newblock \bibinfo{journal}{A\&A} \bibinfo{volume}{545}, \bibinfo{pages}{A113}.
\newblock \DOIprefix\doi{10.1051/0004-6361/201219173}.
\bibitem[{Pushkarev et~al.(2009)Pushkarev, Kovalev, Lister and
  Savolainen}]{pushkarev_jet_2009}
\bibinfo{author}{Pushkarev, A.B.}, \bibinfo{author}{Kovalev, Y.Y.},
  \bibinfo{author}{Lister, M.L.}, et~al.
\newblock , \bibinfo{year}{2009}.
\newblock \bibinfo{journal}{A\&A} \bibinfo{volume}{507},
  \bibinfo{pages}{L33--L36}.
\newblock \DOIprefix\doi{10.1051/0004-6361/200913422}.
\bibitem[{{Pushkarev} et~al.(2017){Pushkarev}, {Kovalev}, {Lister} and
  {Savolainen}}]{pushkarev2017}
\bibinfo{author}{{Pushkarev}, A.B.}, \bibinfo{author}{{Kovalev}, Y.Y.},
  \bibinfo{author}{{Lister}, M.L.}, et~al.
\newblock , \bibinfo{year}{2017}.
\newblock \bibinfo{journal}{MNRAS} \bibinfo{volume}{468},
  \bibinfo{pages}{4992--5003}.
\newblock \DOIprefix\doi{10.1093/MNRAS/stx854}.
\bibitem[{{Sahayanathan} and {Godambe}(2012)}]{sah2012}
\bibinfo{author}{{Sahayanathan}, S.}, \bibinfo{author}{{Godambe}, S.}
\newblock , \bibinfo{year}{2012}.
\newblock \bibinfo{journal}{MNRAS} \bibinfo{volume}{419},
  \bibinfo{pages}{1660--1666}.
\newblock \DOIprefix\doi{10.1111/j.1365-2966.2011.19829.x}.
\bibitem[{Sikora et~al.(2009)Sikora, Stawarz, Moderski, Nalewajko and
  Madejski}]{sikora_constraining_2009}
\bibinfo{author}{Sikora, M.}, \bibinfo{author}{Stawarz, L.},
  \bibinfo{author}{Moderski, R.}, et~al.
\newblock , \bibinfo{year}{2009}.
\newblock \bibinfo{journal}{ApJ} \bibinfo{volume}{704},
  \bibinfo{pages}{38--50}.
\bibitem[{{Tchekhovskoy} et~al.(2008){Tchekhovskoy}, {McKinney} and
  {Narayan}}]{2008MNRAS.388..551T}
\bibinfo{author}{{Tchekhovskoy}, A.}, \bibinfo{author}{{McKinney}, J.C.},
  \bibinfo{author}{{Narayan}, R.}
\newblock , \bibinfo{year}{2008}.
\newblock \bibinfo{journal}{MNRAS} \bibinfo{volume}{388},
  \bibinfo{pages}{551--572}.
\newblock \DOIprefix\doi{10.1111/j.1365-2966.2008.13425.x}.
\bibitem[{Tchekhovskoy et~al.(2009)Tchekhovskoy, McKinney and
  Narayan}]{Tchekhovskoy_2009}
\bibinfo{author}{Tchekhovskoy, A.}, \bibinfo{author}{McKinney, J.C.},
  \bibinfo{author}{Narayan, R.}
\newblock , \bibinfo{year}{2009}.
\newblock \bibinfo{journal}{ApJ} \bibinfo{volume}{699}, \bibinfo{pages}{1789}.
\newblock \DOIprefix\doi{10.1088/0004-637X/699/2/1789}.
\bibitem[{{Thevenet} et~al.(2026){Thevenet}, {Zech}, {Boisson} and
  {Dmytriiev}}]{2026arXiv260205601T}
\bibinfo{author}{{Thevenet}, P.}, \bibinfo{author}{{Zech}, A.},
  \bibinfo{author}{{Boisson}, C.}, et~al.
\newblock , \bibinfo{year}{2026}.
\newblock \bibinfo{journal}{arXiv e-prints} ,
  \bibinfo{pages}{arXiv:2602.05601}\DOIprefix\doi{10.48550/arXiv.2602.05601}.
\bibitem[{{Toscano} et~al.(2025){Toscano}, {G{\'o}mez}, {Zhao}, {Lico},
  {Fuentes}, {Savolainen}, {R{\"o}der}, {Wielgus}, {Pushkarev}, {Traianou},
  {Zeng}, {Gurvits}, {Kovalev}, {P{\"o}tzl} and {Lisakov}}]{Toscano2025}
\bibinfo{author}{{Toscano}, T.}, \bibinfo{author}{{G{\'o}mez}, J.L.},
  \bibinfo{author}{{Zhao}, G.Y.}, et~al.
\newblock , \bibinfo{year}{2025}.
\newblock \bibinfo{journal}{A\&A} \bibinfo{volume}{704}, \bibinfo{pages}{A225}.
\newblock \DOIprefix\doi{10.1051/0004-6361/202555188}.
\bibitem[{{Tramacere} et~al.(2011){Tramacere}, {Massaro} and
  {Taylor}}]{tramacere2011}
\bibinfo{author}{{Tramacere}, A.}, \bibinfo{author}{{Massaro}, E.},
  \bibinfo{author}{{Taylor}, A.M.}
\newblock , \bibinfo{year}{2011}.
\newblock \bibinfo{journal}{ApJ} \bibinfo{volume}{739}, \bibinfo{pages}{66}.
\newblock \DOIprefix\doi{10.1088/0004-637X/739/2/66}.
\bibitem[{{Tramacere} et~al.(2022){Tramacere}, {Sliusar}, {Walter}, {Jurysek}
  and {Balbo}}]{tramacere2022}
\bibinfo{author}{{Tramacere}, A.}, \bibinfo{author}{{Sliusar}, V.},
  \bibinfo{author}{{Walter}, R.}, et~al.
\newblock , \bibinfo{year}{2022}.
\newblock \bibinfo{journal}{A\&A} \bibinfo{volume}{658}, \bibinfo{pages}{A173}.
\newblock \DOIprefix\doi{10.1051/0004-6361/202142003}.
\bibitem[{{Yan} et~al.(2025){Yan}, {Cui}, {Hada}, {Frey}, {Lu}, {Chen}, {Xu},
  {Fariyanto} and {Ho}}]{2025ApJ...991...75Y}
\bibinfo{author}{{Yan}, X.}, \bibinfo{author}{{Cui}, L.},
  \bibinfo{author}{{Hada}, K.}, et~al.
\newblock , \bibinfo{year}{2025}.
\newblock \bibinfo{journal}{ApJ} \bibinfo{volume}{991}, \bibinfo{pages}{75}.
\newblock \DOIprefix\doi{10.3847/1538-4357/adf84c}.
\bibitem[{{Yi} et~al.(2024){Yi}, {Park}, {Nakamura}, {Hada} and
  {Trippe}}]{acceleration_obs}
\bibinfo{author}{{Yi}, K.}, \bibinfo{author}{{Park}, J.},
  \bibinfo{author}{{Nakamura}, M.}, et~al.
\newblock , \bibinfo{year}{2024}.
\newblock \bibinfo{journal}{A\&A} \bibinfo{volume}{688}, \bibinfo{pages}{A94}.
\newblock \DOIprefix\doi{10.1051/0004-6361/202449790}.
\bibitem[{{Zacharias}(2023)}]{Zacharias2023}
\bibinfo{author}{{Zacharias}, M.}
\newblock , \bibinfo{year}{2023}.
\newblock \bibinfo{journal}{A\&A} \bibinfo{volume}{669}, \bibinfo{pages}{A151}.
\newblock \DOIprefix\doi{10.1051/0004-6361/202244683}.
\bibitem[{{Zakamska} et~al.(2008){Zakamska}, {Begelman} and
  {Blandford}}]{2008ApJ...679..990Z}
\bibinfo{author}{{Zakamska}, N.L.}, \bibinfo{author}{{Begelman}, M.C.},
  \bibinfo{author}{{Blandford}, R.D.}
\newblock , \bibinfo{year}{2008}.
\newblock \bibinfo{journal}{ApJ} \bibinfo{volume}{679},
  \bibinfo{pages}{990--999}.
\newblock \DOIprefix\doi{10.1086/587870}.
\bibitem[{{Zhang} \& {Giannios}(2021)}]{Zhang_2021}
\bibinfo{author}{{Zhang}, H.}, \bibinfo{author}{{Giannios}, D.}
\newblock , \bibinfo{year}{2021}.
\newblock \bibinfo{journal}{MNRAS} \bibinfo{volume}{502},
  \bibinfo{pages}{1145--1157}.
\newblock \DOIprefix\doi{10.1093/mnras/stab008}.

\end{thebibliography}

\newpage 

\appendix

\section{BLR $\gamma\gamma$ absorption}
\label{appendixA}

\begin{figure}[h!]
\resizebox{\hsize}{!}{\includegraphics[clip=true]{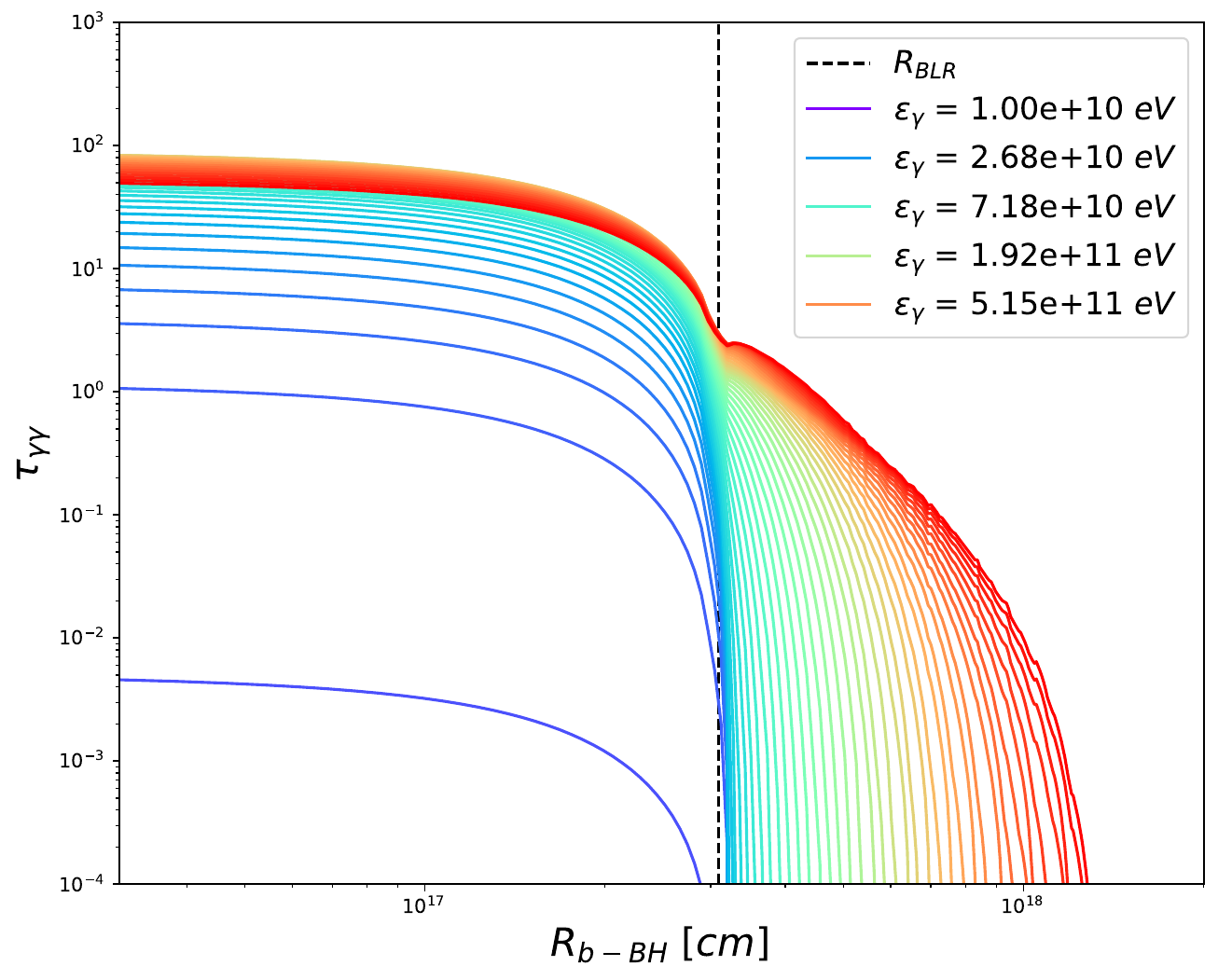}}
\caption{$\gamma\gamma$ absorption optical depth as a function of the distance between the black hole and the emitting region $R_{b-BH}$ and of the energy of the emitted photon in the BLR frame $\epsilon_\gamma$.}
\label{fig:taugg}
\end{figure}

The very high energy emission coming from the blob is expected to be partially absorbed by the BLR radiation field through $\gamma\gamma$ absorption. We model this absorption as described in \citet{Bottcher_2016}, adopting a thin BLR shell (width of the BLR $\sim 0.1 R_{\mathrm{BLR}}$) to correspond to our model and using Eq.\ref{rad_BLR}, converted into the BLR rest frame, to describe the BLR radiation field.

The resulting $\gamma\gamma$ absorption optical depth $\tau_{\gamma\gamma}$ as a function of the energy of the emitted photon and of the distance between the black hole and the emitting region (the blob) is plotted in Fig.\ref{fig:taugg}. As can be seen, the $\gamma\gamma$ absorption is not expected to be relevant for emitted photons of energy below $10 \ \mathrm{GeV}$. 

The $\gamma\gamma$ absorption from the disk and dusty torus radiation fields is not included in our model because it is significant for energies $>1 \ \mathrm{TeV}$, which is two orders of magnitude greater than the observed flux points from {\it Fermi}-LAT.
Absorption on the dusty torus could be relevant in the case of a potential VHE flare at energies above $100$ GeV (Fig.~\ref{fig:lcdata}). But from Fig.~\ref{fig:Uext} one can estimate that the optical depth for the torus is roughly $U'_\mathrm{BLR}/U'_\mathrm{DT}$ lower than the optical depth for the BLR and thus negligible.  

\section{Adiabatic expansion of propagating plasma blobs}
\label{appendixB}
As a blob moves within a jet with non-zero opening angle, it undergoes adiabatic expansion. Following \citet{tramacere2022}, we describe the evolution of the radius of a blob, in the blob frame, as such:
\begin{equation}
    R(t) = R_0 + \beta_{\mathrm{exp}} c t,
    \label{r_exp}
\end{equation}
with $R_0$ its initial radius and $\beta_{exp}$ the reduced expansion velocity. The latter is deduced from the geometry of the jet. 
\begin{equation}
    \beta_{\mathrm{exp}} = \beta_{\mathrm{blob}}\tan(\theta_o),
    \label{beta_exp}
\end{equation}
with $\theta_o$ the jet intrinsic opening angle and $\beta_{\mathrm{blob}}$ the blob reduced velocity. The intrinsic opening angle is computed as function of the Lorentz factor of the blob (and of the jet):
\begin{equation}
    \theta_0 = \alpha / \Gamma_{b-BH},
\end{equation}
with the best fitted value found by \citet{pushkarev_jet_2009} $\alpha = 0.26$. The intrinsic opening angle is thus evolving with time following Eq.~\ref{acc_blob}.

Because of flux freezing and energy conservation \citep{RevModPhys.56.255}, the magnetic field is evolving with time:
\begin{equation}
    B(t) = B_0\left(\frac{R_0}{R(t)}\right)^{m_B},
\end{equation}
with $m_B\in[1,2]$ an index dependent of the geometric configuration of the magnetic field. In our model it is set to $m_B =1$ but several tests showed that the variation of this parameter does not have a significant impact on the presented results.

\section{Electron distributions}
\label{appendixC}

\begin{figure}[h!]
\resizebox{\hsize}{!}{\includegraphics[clip=true]{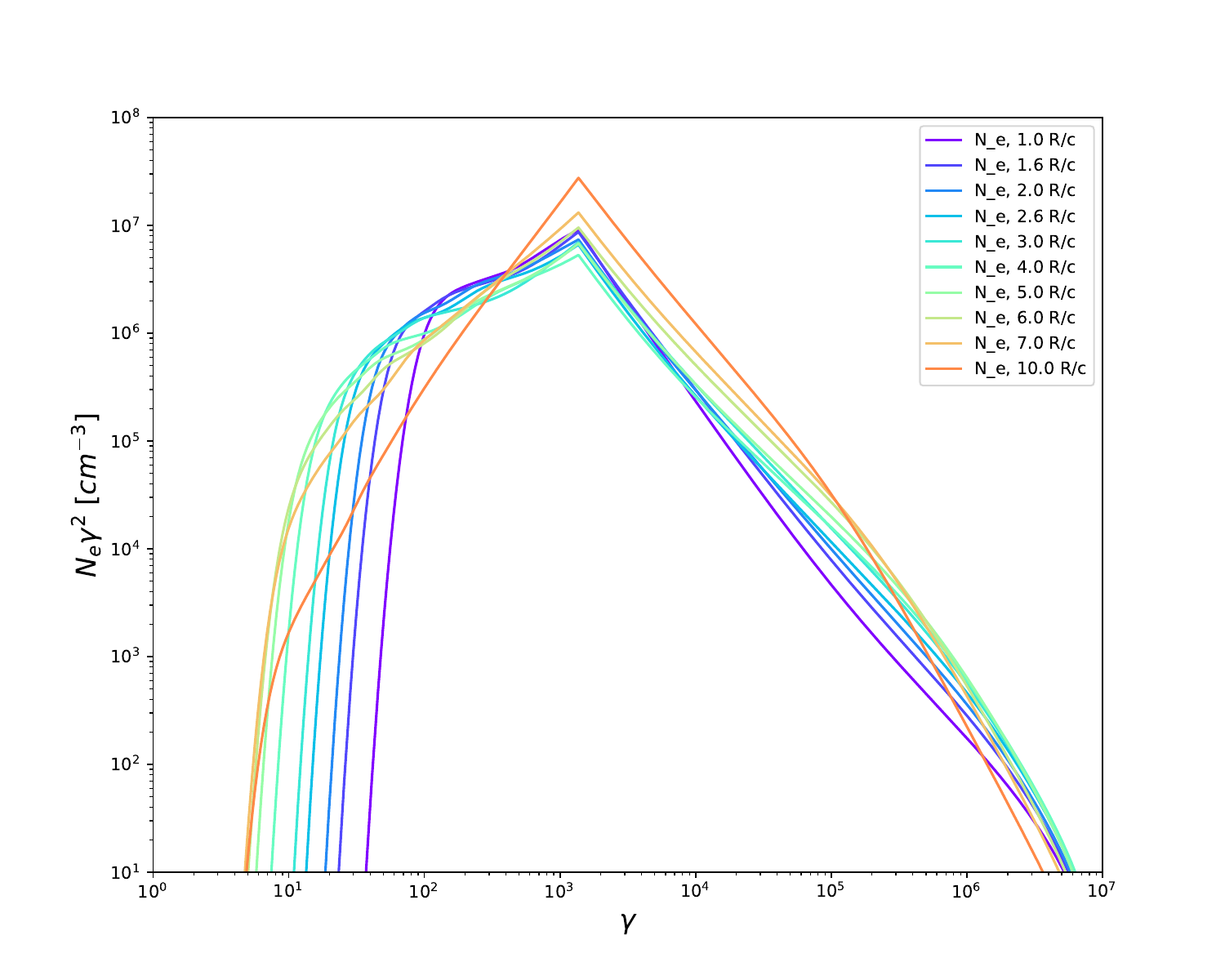}}
\caption{Electron distribution of blob 2 during its early evolution, crossing the BLR at $\approx 5.0 \, R/c$.}
\label{fig:ED1}
\end{figure}

\begin{figure}[h!]
\resizebox{\hsize}{!}{\includegraphics[clip=true]{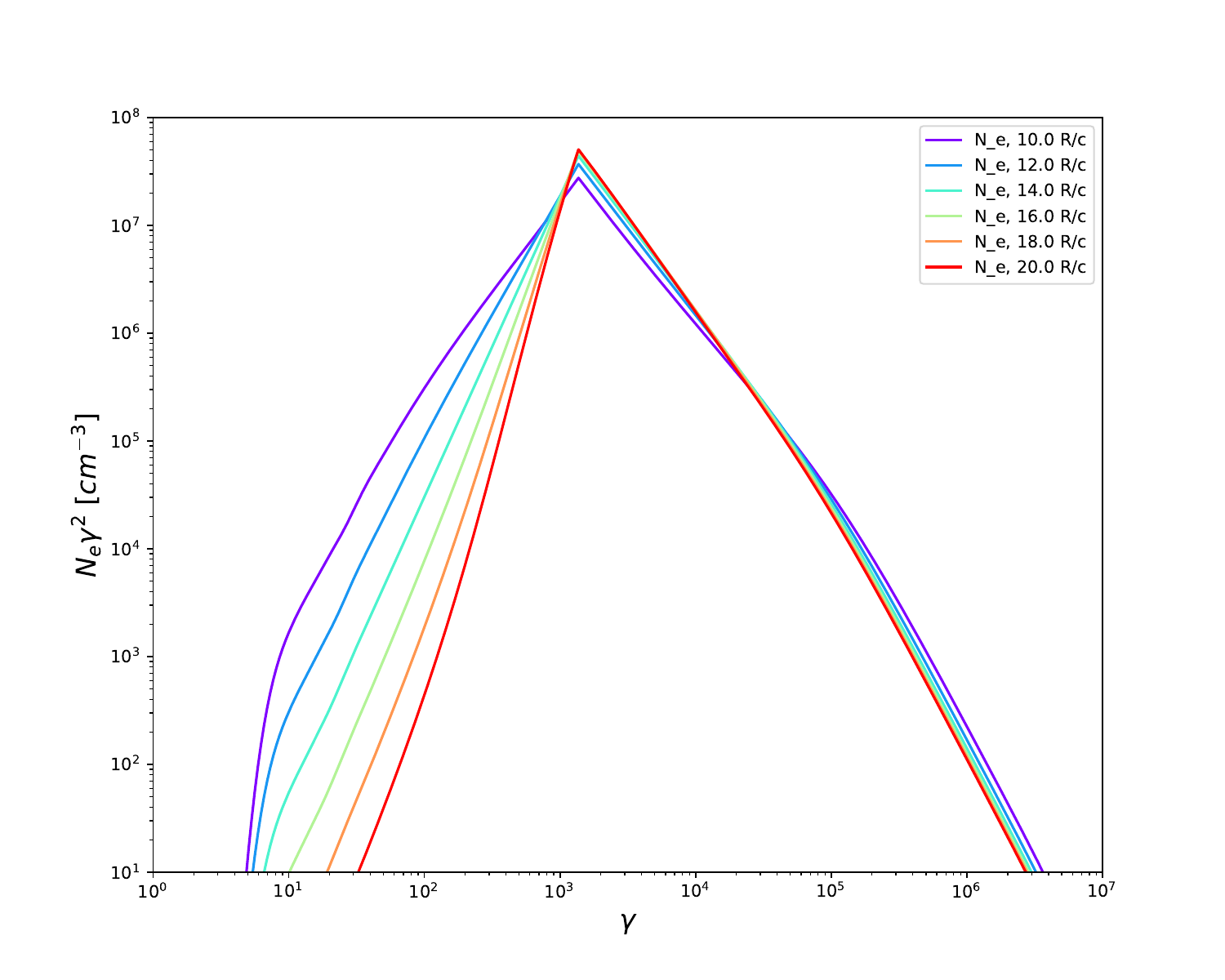}}
\caption{Electron distribution of blob 2 after crossing the BLR, between $10$ and $20 \, R/c$.}
\label{fig:ED2}
\end{figure}

\begin{figure}[h!]
\resizebox{\hsize}{!}{\includegraphics[clip=true]{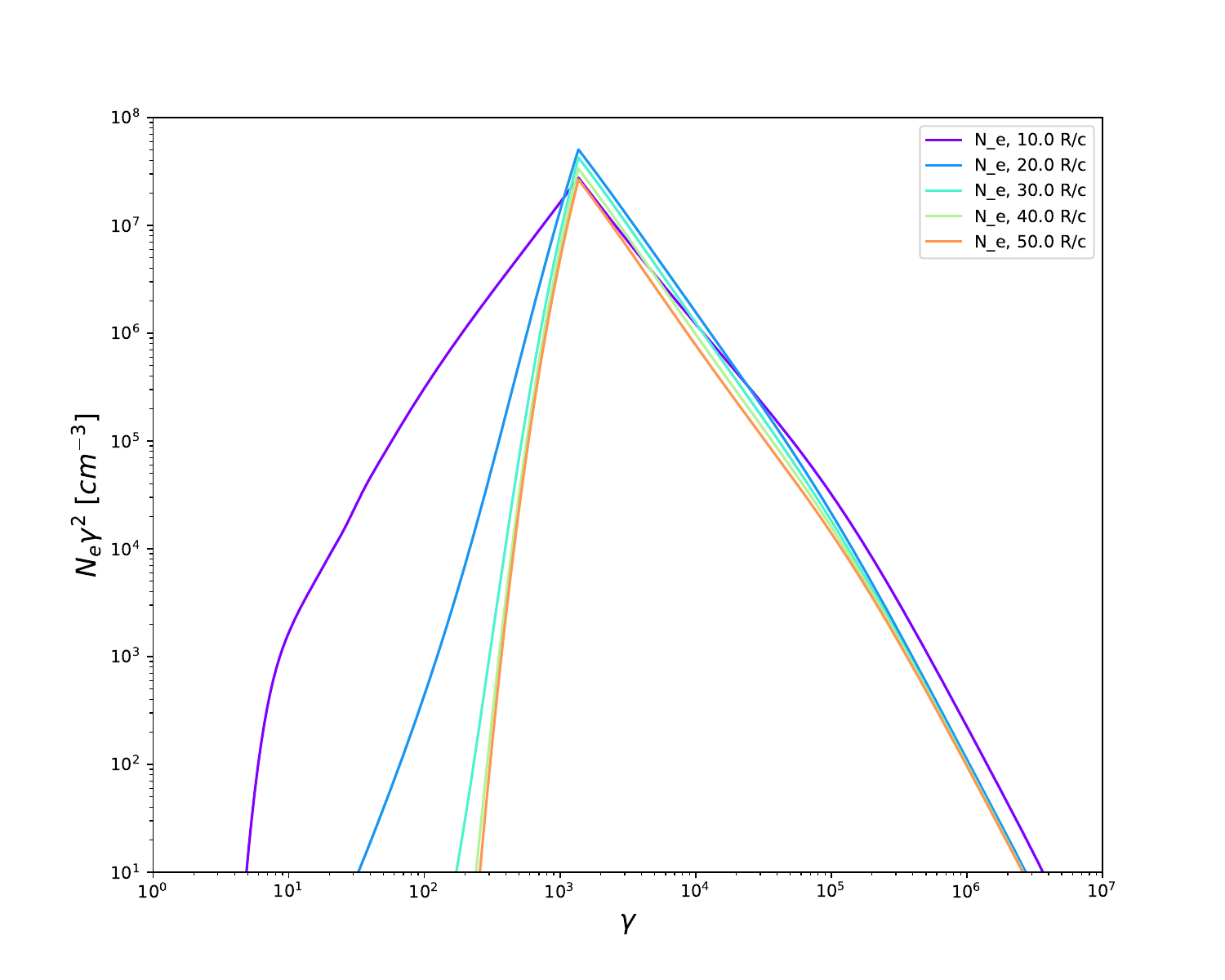}}
\caption{Electron distribution of blob 2 during its late stage, between $10$ and $50 \, R/c$.}
\label{fig:ED3}
\end{figure}

\begin{figure}[h!]
\resizebox{\hsize}{!}{\includegraphics[clip=true]{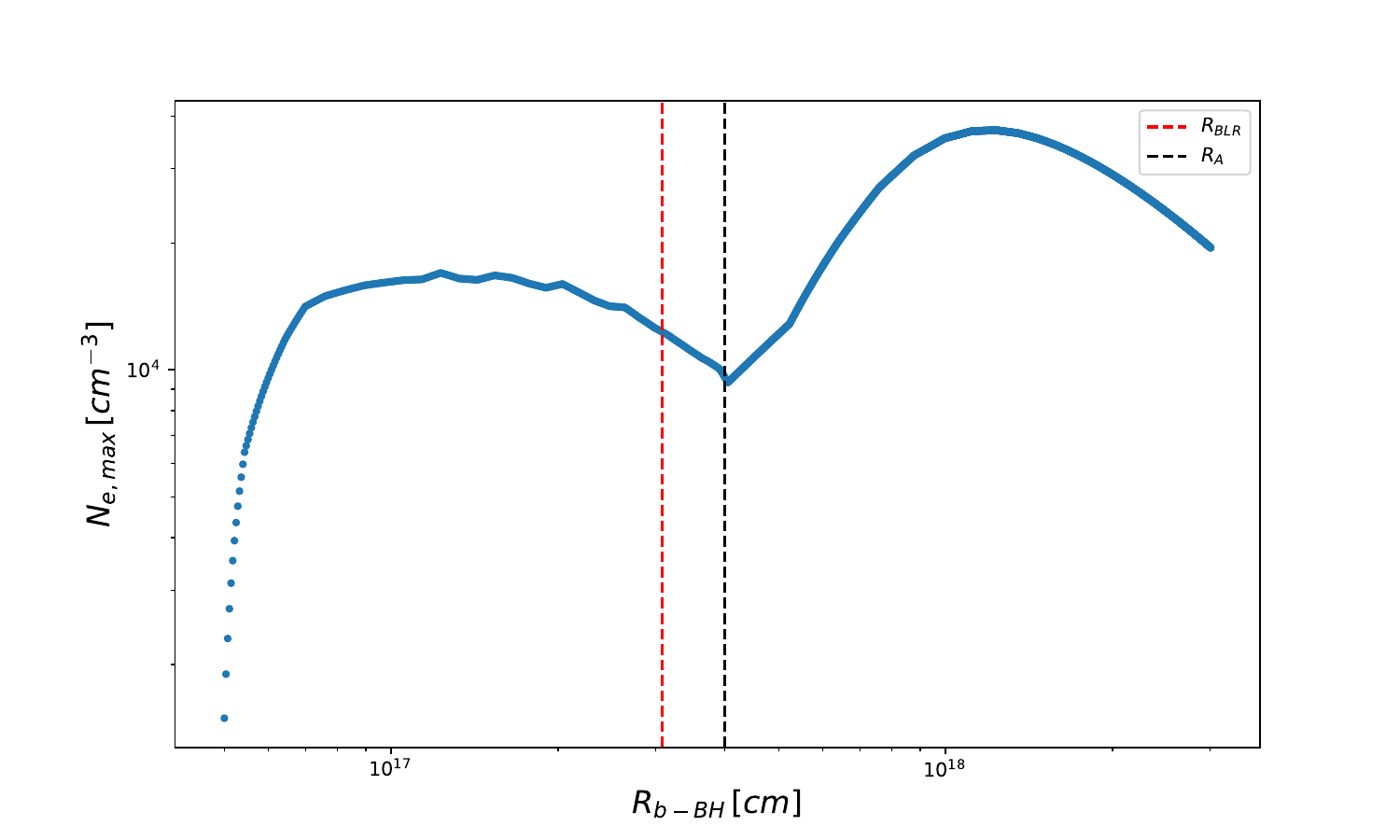}}
\caption{Maximum of the electron distribution of blob 2 as a function of distance to the black hole.}
\label{fig:ED4}
\end{figure}

This section details the evolution of the electron distribution of blob~2 throughout its propagation. 
During the initial phase ($t < 5 \, R/c$), as the blob approaches the BLR, 
the electron distribution is attenuated by EIC cooling. 
Upon crossing the BLR ($t > 5 \, R/c$), the distribution is enhanced due to the 
abrupt decline in the external photon field density. Finally, at late 
evolutionary stages ($t > 20 \, R/c$), the electron distribution decreases 
again, driven by adiabatic expansion. This can be seen in Fig.~\ref{fig:ED1} to~\ref{fig:ED3}. The maximum of the electron distribution is plotted in Fig.~\ref{fig:ED4}, as a function of distance.

\end{document}